# Simulation of Coupled Heat and Mass Transport With Reaction in PEM Fuel Cell Cathode using Lattice Boltzmann Method

**Jithin M**
Department of Mechanical Engineering,
Indian Institute of Technology Kanpur,
Kanpur, Uttar Pradesh, India.
E-mail: jithinm@iitk.ac.in

**Saurabh Siddharth**
Department of Mechanical Engineering,
Indian Institute of Technology Kanpur,
Kanpur, Uttar Pradesh, India.
E-mail: sausid@iitk.ac.in

**Malay K Das**
Department of Mechanical Engineering,
Indian Institute of Technology Kanpur,
Kanpur, Uttar Pradesh, India.
E-mail: mkdas@iitk.ac.in

**Ashoke De[1]**
Department of Aerospace Engineering,
Indian Institute of Technology Kanpur,
Kanpur, Uttar Pradesh, India.
E-mail: ashoke@iitk.ac.in
Tel.: +91-512-6797863; Fax: +91-512-6797561

## ABSTRACT

*Fluid, heat and species transport and oxygen reduction in the cathode of a PEM fuel cell are simulated using multi relaxation lattice Boltzmann method. Heat generation due to oxygen reduction and its effects on transport and reaction are considered. Simulations for various cell operating voltages, temperatures and flow rates with various values of porous media properties, namely, permeability, porosity, and effective porous media diffusion coefficient, are performed to study transport and operating characteristics in the electrode. It is seen that maximum output power density achievable is limited by the*

---
[1] Corresponding author

*mass transport rate. A small increase in current density is obtained by increasing the operating temperature. However, this results in an increase in the rate of heat generation. Permeability and porosity of the gas diffusion layer do not show a significant impact on the performance in the range of values presently simulated. Higher permeability, in turn, resulted in enhancement of thermal gradients in the porous layer. A significant increase in the maximum current density obtainable is observed with increase in flow rates. The higher convection associated with high flow rate facilitates better transport of species and heat at the catalyst layer resulting in larger current density with a lesser chance of hotspot formation. Increased species diffusion coefficient also resulted in increasing the power output substantially. In addition, the fuel utilization is also improved at high diffusion rates in the porous media. The study analyses and shows the impact of various operating and material parameters affecting the performance of a PEM fuel cell with special attention on enhancing the maximum power density attainable.*

## 1. INTRODUCTION

The range of efficient and safe operation of different fuel cell types varies from the high temperatures of Solid Oxide Fuel Cells ($800^0C$ - $1000^0C$) and Molten Carbonate Fuel Cells ($600^0C$ - $700^0C$) to that of Polymer Electrolyte Membrane Fuel Cells ($70^0C$ - $90^0C$), based on the operation, design and material properties. Among these, the polymer electrolyte membrane (PEM) fuel cells are becoming quite popular in many practical applications owing to their advantages like high efficiency, zero pollution, and low operating temperatures. But their performance is strongly dependent on the temperature and water content at the membrane electrode assembly (MEA) since phenomena like electrochemical reaction, transport properties, and proton conductivity are highly dependent on the water content and local temperature in the MEA.

The control of water content in the PEM fuel cell in operation is required to ensure optimum hydration level in the electrolyte membrane to facilitate proton conduction and should also ensure liquid water removal from the porous electrodes to prevent flooding of the pores. Likewise, thermal management involves the control of operating temperature from being so low that the efficiency is affected and also the iso-thermalization of its components to prevent hot spot formations. Thermal management of fuel cells using heat transfer devices like heat pipes have been studied in the past and the results seem to be very encouraging. Heat pipes have been put to use in enhancing heat transfer in fuel cell stacks varying from microscale to the scale of stationary power generation and in other ancillary systems [1]. In addition to selection and design of efficient heat transfer devices, the knowledge of the heat and water transport mechanism occurring at the MEA under various design and operating conditions is necessary.

Numerous models for the numerical study of transport phenomena in PEMFC have been proposed in the literature which helps to understand the transport of the different components of PEMFC. Early works involving the computational study of PEMFCs include those by Springer, Zawodzinski [2], Bernardi and Verbrugge [3], Nguyen and White [4] and Fuller and Newman [5]. An entire fuel cell sandwich was modeled including the solution of flow, temperature distribution, and species concentration distribution by Gurau, Liu [6] as early as 1998. More recently, Dutta [7] carried out simulations of straight channel PEM fuel cells including an anode, cathode, and membrane. Three-dimensional flow solver FLUENT was used in their work to solve for flow, phase change and electrochemical reactions in the domain. Later in 2005, Ju, Wang [8] conducted 3D non-isothermal single phase modeling of PEMFC using commercial CFD package STAR-CD. Multiphase non-isothermal modeling was conducted by Birgersson, Noponen [9] in 2005 using finite element solver FEMLAB 2.3. A three-dimensional comprehensive model for a PEMFC unit cell was developed by Sui et al. [10, 11] using a commercial CFD solver, CFD-ACE+. Later in 2013, a combined thermal and water management was studied from simulations of three-dimensional non-isothermal multi phase simulations of PEMFC using Finite Volume Method by Cao, Lin [12]. Very recently, Xing, Liu [13] presented a two-dimensional multi phase non-isothermal agglomerate model for PEMFC using commercial software COMSOL Multiphysics 2.3a. A review of the fundamental models of PEMFC was given by Wang [14] in 2004. A detailed review of literature of numerical works conducted in PEMFCs is available from Bednarek and Tsotridis [15] also.

In addition to the works mentioned above, numerous works have been performed in the computational simulation in PEMFCs using Lattice Boltzmann Method (LBM). LBM is gaining much popularity as a promising method due to its advantages including easy parallelization, ease in handling complex geometry and boundary conditions, the capability to capture phenomena in small length scales etc. Recent studies [16, 17] have proved that LBM offers clear advantages over conventional CFD techniques in the macroscale simulations also. According to the literature [16], the major advantages of LBM, compared to conventional CFD techniques, include (a) linear convection term with a relaxation process to mimic the macroscopic equations, (b) pressure obtained from an equation of state, in contrast to the Poisson equation solution and (c) finite number of velocities in the phase space, making calculation of macroscopic quantities, from microscopic distribution function, simple. In addition, it is shown in the literature that LBM is computationally more efficient than finite difference method in a solution of flow in porous media [18] and needs less time to produce results for the same grid size. According to Chai, Guo [19] even if LBM require finer grids to produce results of the same accuracy as other conventional CFD methods, it provides flexibility, efficiency, easy implementation of boundary conditions and amenability towards parallelization while handling complex flows.

However, according to Wang and Afsharpoya [20], the accuracy and reliability of the Lattice-Boltzmann models and the implementation issues associated with practical applications, such as non-uniform grid, forcing implementation, boundary conditions, and porous-medium interface etc. have to be studied in detail before its reliable application. They attempted a study of the implementation details of LBM by simulating flow through serpentine flow channels and channels partially filled with porous media. In a very recent publication by Molaeimanesh, Googarchin [21], the works using LBM in PEMFC simulation has been reviewed. In this article, LBM works on PEMFCs are classified based on the purpose of the study as (1) evaluation of transport properties, (2) water droplet dynamics and (3) operation analysis. LBM for evaluation of properties involves simulation of flow through the porous microstructure of the porous components of the fuel cell [22-33]. The detailed geometry of the microstructure is usually reconstructed from experimental imaging. Results from such simulations provide information necessary for macroscale modeling, like values of permeability, diffusivity, conductivity, relative permeability relations etc. In the second type of works mentioned, the transport of liquid water formed from electrochemical reactions is studied. This includes multiphase flow simulations inside the porous gas diffusion layer (GDL)[34-38] and the droplet formation, detachment and transport in the flow channel[39, 40]. This study is particularly important in ensuring proper water management and reactant transport to the reaction sites. Studies have shown that the gas flow velocity and the GDL surface wettability plays the leading role in the droplet dynamics and hence the water removal from the electrodes. Some studies also studied the gas channel geometry effect on water transport [41-43]. The third class, namely the electrode operation analysis involves the transport of the species, electric potential and heat including the electrochemistry [35, 44-49]. The works in this class can further be subdivided based on the catalyst layer model chosen into thin film models, discrete catalyst volume models and agglomerate models. From the simulations the electrochemical reaction rates are obtained which give the details of the operation of the cell in the form of operating voltage and the current density. Performance curves are plotted from these quantities and the parameter affecting it are analyzed.

The present work falls in the third category in which the performance of the cathode of a PEMFC is analysed from simulations of fluid flow, temperature, and species transport using multi relaxation time (MRT) LBM in a geometry resembling the cathode of a PEM fuel cell with exothermic electrochemical reaction taking place at the catalyst layer. The catalyst layer is assumed to be of negligible thickness. The products and the heat generated by the reaction are removed through the flow outlet. The increase in temperature affects the reaction rate and the species transport. Attention is focused on the reactant transport under single phase conditions through the cathode GDL under different operating conditions and the effect of reactant transport rate on the performance is studied. Interest is primarily focussed on the factors affecting the reactant transport to the reaction sites so that the maximum obtainable output power density can be enhanced. Detailed parametric study and analysis of results are conducted on the various operation and material factors affecting performance.

The outline of the remaining part of the paper is as follows. Section 2 discusses the solution of fluid flow, mass transport and heat transport equations, conducted using FORTRAN code developed in-house using separate distribution functions for each equation. The details of implementation can be found in Guo and Zhao [50]. Section 3 includes the validation of the code conducted for (a) flow and heat transfer in clear media, (b) flow and heat transfer in porous media and (c) heterogeneous reaction on a solid surface. Section 4 discusses the computational domain and boundary conditions along with the reaction kinetics and transport coefficient calculations. Section 4 is the results and discussions where the importance of reactant mass transport and heat transport in the porous GDL is discussed as the operating voltage is varied. Then the influence of operating temperature and reactant flow rates on performance are observed. The porous material properties namely, the porosity, permeability and porous media effective species transport coefficient are varied and their impact on cell life and performance are discussed. Finally, in section 6, the major conclusions from the study are summarized.

## 2. LATTICE BOLTZMANN FORMULATION

### 2.1. Fluid Flow

In Lattice Boltzmann Method, transport of an ensemble of molecules through a regular lattice is solved to obtain the macroscopic parameters. The number of molecules at each lattice point at an instant of time, moving in a particular discretized direction is obtained from the particle density distribution function $\mathbf{f}(\mathbf{x},t)$, from the equation:

$$\mathbf{f}(\mathbf{x}+\mathbf{e}\Delta t, t+\Delta t) - \mathbf{f}(\mathbf{x},t) = -\mathbf{M}^{-1}\mathbf{S}\left[\mathbf{m}(\mathbf{x},t) - \mathbf{m}^{eq}(\mathbf{x},t)\right] + \Delta t \mathbf{M}^{-1}(\mathbf{I}-\mathbf{S}/2)\mathbf{G} \qquad (1)$$

The LHS is the streaming part in which $\mathbf{f}(\mathbf{x},t)$ gets streamed to $\mathbf{f}(\mathbf{x}+\mathbf{e}\Delta t, t+\Delta t)$ at time $t+\Delta t$. $\mathbf{e}$ is the discretized lattice velocity vector. In the present work, a D2Q9 model is selected with the velocity being discretized into nine components in a two-dimensional space. The scalar components of the velocity vector $\mathbf{e}$ are defined as:

$$e_i = \begin{cases} (0,0) & i=0 \\ c\left[\cos\left(\dfrac{i-1}{2}\pi\right), \sin\left(\dfrac{i-1}{2}\pi\right)\right] & i=1,2,3,4 \\ \sqrt{2}c\left[\cos\left(\dfrac{i-5}{2}\pi+\dfrac{\pi}{4}\right), \sin\left(\dfrac{i-5}{2}\pi+\dfrac{\pi}{4}\right)\right] & i=5,6,7,8 \end{cases} \qquad (2)$$

Here $c = \Delta x / \Delta t$ is the lattice speed and is taken as 1 in lattice units. The first term in the RHS of Eq. (1) is the collision term done in the moment space, as the transformation matrix $\mathbf{M}$ maps the distribution function $\mathbf{f}$ from the velocity space to the moment space. The transformation matrix $\mathbf{M}$ for D2Q9 model is

$$M = \begin{bmatrix} 1 & 1 & 1 & 1 & 1 & 1 & 1 & 1 & 1 \\ -4 & -1 & -1 & -1 & -1 & 2 & 2 & 2 & 2 \\ 4 & -2 & -2 & -2 & -2 & 1 & 1 & 1 & 1 \\ 0 & 1 & 0 & -1 & 0 & 1 & -1 & -1 & 1 \\ 0 & -2 & 0 & 2 & 0 & 1 & -1 & -1 & 1 \\ 0 & 0 & 1 & 0 & -1 & 1 & 1 & -1 & -1 \\ 0 & 0 & -2 & 0 & 2 & 1 & 1 & -1 & -1 \\ 0 & 1 & -1 & 1 & -1 & 0 & 0 & 0 & 0 \\ 0 & 0 & 0 & 0 & 0 & 1 & -1 & 1 & -1 \end{bmatrix} \qquad (3)$$

The products $\boldsymbol{m} = \boldsymbol{Mf}$ and $\boldsymbol{m}^{eq} = \boldsymbol{Mf}^{eq}$ are vectors in the moment space, and the nine velocity moments are:

$$\mathbf{m} = \left( \rho, e, \zeta, j_x - \frac{\Delta t}{2}\rho F_x, q_x, j_y - \frac{\Delta t}{2}\rho F_y, q_y, p_{xx}, p_{xy} \right)^{\mathrm{T}} \qquad (4)$$

Here $\rho$ is the fluid density, $\zeta$ is related to the energy and $j_x - \rho F_x \Delta t / 2$ is related to the square of energy and $j_y - \Delta t \rho F_y / 2$ are the components of momentum in x and y directions. $q_x$ and $q_y$ are related to the components of energy flux in x and y directions. $p_{xx}$ and $p_{xy}$ correspond to the symmetric and traceless parts of the strain rate tensor. The components of equilibrium moments for flow through porous media are given by:

$$\mathbf{m}^{eq} = \left( \rho_0, e^{eq}, \zeta^{eq}, j_x, q_x^{eq}, j_y, q_y^{eq}, p_{xx}^{eq}, p_{xy}^{eq} \right)^{\mathrm{T}}; \rho_0 = \text{mean fluid density} = 1 \text{ in lattice unit}$$
$$e^{eq} = -4\rho_0 + 6\phi p + 3\phi^{-1}\rho_0 |\mathbf{u}|^2; \zeta^{eq} = 4\rho_0 - 9\phi p - 3\phi^{-1}\rho_0 |\mathbf{u}|^2; j_x = \rho_0 u_x; j_y = \rho_0 u_y \qquad (5)$$
$$q_x^{eq} = -\rho_0 u_x; q_y^{eq} = -\rho_0 u_y; p_{xx}^{eq} = \rho_0 \phi^{-1}\left( u_x^2 - u_y^2 \right); p_{xy}^{eq} = \rho_0 \phi^{-1} u_x u_y$$

One of the advantages of the MRT collision is that separate relaxation times can be assigned to each moment. Hence the relaxation matrix, $\mathbf{S}$, is a diagonal matrix with relaxation times for each moment as the diagonal elements. The nine diagonal elements as given in [51, 52] are $0, \dfrac{1}{\tau}, \dfrac{1}{\tau}, 0, 8\left(\dfrac{2-1/\tau}{8-1/\tau}\right), 0, 8\left(\dfrac{2-1/\tau}{8-1/\tau}\right), \dfrac{1}{\tau}, \dfrac{1}{\tau}$

The second term of Eq. (1) is the body force term which arises due to the presence of porous media. The components of the forcing vector $G$ in the moment space are:

$$\begin{aligned}
\mathbf{G} &= \left(G_0, G_1, G_2, G_3, G_4, G_5, G_6, G_7, G_8\right)^{\mathrm{T}} \\
G_0 &= 0; G_1 = 6\rho_0 \phi^{-1} \mathbf{u} \cdot \mathbf{F}; G_2 = -6\rho_0 \phi^{-1} \mathbf{u} \cdot \mathbf{F}; G_3 = \rho_0 F_x; G_4 = -\rho_0 F_x; G_5 = \rho_0 F_y \\
G_6 &= -\rho_0 F_y; G_7 = -2\rho_0 \phi^{-1}\left(u_x F_x - u_y F_y\right); G_8 = \rho_0 \phi^{-1}\left(u_x F_x + u_y F_y\right)
\end{aligned} \qquad (6)$$

Here $\mathbf{F}$ is the sum of the Darcy and Forchheimer force terms given by:

$$\mathbf{F} = -\phi v K^{-1}\mathbf{u} - \phi F_\phi K^{-.5}|\mathbf{u}|\mathbf{u}; F_\phi = .143\phi^{-1.5}; K = .0067\phi^3(1-\phi)^{-2}d_P^2 \qquad (7)$$

A solution of Eq. (1) gives the distribution function from which the macroscopic velocity, pressure, and kinematic viscosity can be obtained as follows:

$$\mathbf{u} = \frac{\mathbf{v}}{l_0 + \sqrt{l_0^2 + l_1|\mathbf{v}|}}$$

$$\mathbf{v} = \sum_{i=0}^{8} \frac{\mathbf{e}_i f_i}{\rho_0}$$

$$l_0 = \frac{2K + \phi v \Delta t}{4K}$$

$$l_1 = \frac{\phi \Delta t F_\phi}{2\sqrt{K}} \tag{8}$$

$$p = \frac{.6}{\phi} \sum_{i=1}^{8} f_i - \frac{6\rho_0 |\mathbf{u}|^2}{15\phi^2}$$

$$\frac{3v}{\Delta t} = \tau - .5$$

## 2.2 Temperature and species transport

D2Q5 velocity discretization model is adopted for temperature and species transport as they are scalar functions. The discretized velocity directions are defined as [16]:

$$\mathbf{e}_i = \begin{cases} (0,0) & i=0 \\ c\left[\cos\left(\frac{i-1}{2}\pi\right), \sin\left(\frac{i-1}{2}\pi\right)\right] & i=1,2,3,4 \end{cases} \tag{9}$$

The lattice Boltzmann equation, transformation matrix, and the equilibrium moment space vector are given by [52]:

$$\mathbf{g}(\mathbf{x}+\mathbf{e}\Delta t, t+\Delta t) - \mathbf{g}(\mathbf{x},t) = -\mathbf{N}^{-1}\mathbf{Q}\left[\mathbf{n}(\mathbf{x},t) - \mathbf{n}^{eq}(\mathbf{x},t)\right]$$

$$\mathbf{n}^{eq} = (T, \mathbf{u}T, \mathbf{v}T, bT, 0)^{\mathrm{T}}; b = 20\sqrt{3}\alpha - 4$$

$$\mathbf{N} = \begin{bmatrix} 1 & 1 & 1 & 1 & 1 \\ 0 & 1 & 0 & -1 & 0 \\ 0 & 0 & 1 & 0 & -1 \\ -4 & 1 & 1 & 1 & 1 \\ 0 & 1 & -1 & 1 & -1 \end{bmatrix}; \mathbf{Q} = diag\left[0, \frac{6}{3+\sqrt{3}}, \frac{6}{3+\sqrt{3}}, \frac{6}{3+2\sqrt{3}}, \frac{6}{3+2\sqrt{3}}\right] \tag{10}$$

Here **g** is the distribution function. Separate distribution functions are solved for temperature and concentration of each species being considered, using the appropriate values of the transport properties. The transformation matrix maps the distribution functions to the space of velocity moments. For species, the temperature will be replaced by the concentration in each of the components. Finally, the relaxation matrix **Q** is a 5×5 diagonal matrix with the diagonal elements shown in the Eq. (10). The temperature or concentration is obtained from the distribution function according to the relation:

$$T = \sum_{i=0}^{4} g_i \qquad (11)$$

*T* in this equation will be temperature for energy equation and concentration for species equation.

## 3. VALIDATION

*Flow and heat transfer in clear media*: For validation of the formulation, natural convection flow in a square cavity with the heated wall is simulated and the results are compared with those of Roy and Basak [53]. The left and bottom walls of the cavity are at a higher temperature while the right wall is at a lower temperature. The top wall is assumed to be adiabatic. Simulations are carried out for different values of Rayleigh and Prandtl numbers and the variation of the local Nusselt number at the bottom wall is plotted as shown in Fig. 1.

*Flow and heat transfer in porous media*: Results of flow and heat transfer in porous media shown in Fig. 2 is compared with Guo and Zhao [50]. Natural convection in a cavity filled porous media is simulated for different values of Rayleigh and Darcy numbers. Figure 2 shows the temperature profile at mid-height of the cavity.

*Surface reaction*: Diffusion and surface reaction with linear kinetics in a rectangular domain is carried out and Fig. 3 shows the results being compared with Kang, Lichtner [54]. The concentration distribution of the component undergoing surface reaction is shown in the figure. The value of Damkohler number is varied which represents the rate of the surface reaction.

All the tests conducted gave results which show good agreement with literature. The code being used has already been extensively validated and was invoked in other works also [55-57].

## 4. COMPUTATIONAL DOMAIN AND BOUNDARY CONDITIONS

The computational domain is shown in the Fig. 4. The Lattice Boltzmann (LB) equations for fluid flow, temperature and concentration are solved in the entire domain with different conditions in clear media, porous media and the reaction surface. The formulation and parameters used for the simulation of electrochemical reaction are adopted from Ju, Wang [8], Gu and Wang [58] and Um, Wang [59]. The schematic in Figure 4 shows a 2D section of the anode and cathode regions of a PEMFC. The computational domain consists of the cathode gas diffusion layer (GDL) which is the porous layer and the flow channel which is clear medium, where porosity, $\phi$, is set to 1. For GDL, $\phi = 0.4$ and permeability, $K = Da \times h^2$. Here 'Da' is the Darcy number and 'h' is the channel height. On all solid surfaces, bounce back condition is applied so that the no-slip condition for velocity and insulated condition for concentration and temperature is satisfied.

Along the catalyst layer the following oxygen reduction reaction takes place:

$$4H^+ + O_2 + 4e^- \rightarrow 2H_2O \qquad (12)$$

The transport of $H^+$ and $e^-$ is not considered. The rate of reaction is dependent on the concentration of the oxygen reaching the surface and the activation overpotential according to the Tafel kinetics is given by

$$j = -a j_{0,c}^{ref} \left( \frac{C_{O_2}}{C_{O_2,ref}} \right) \exp\left( -\frac{\alpha_c F}{RT} \eta \right) \qquad (13)$$

The over-potential value depends on the operating voltage as

$$\eta = V_{op} - U_0 \qquad (14)$$

The reaction causes heat generation due to the irreversible heat of electrochemical reaction and the reversible entropic heat. Joule heating is not considered in the present work. The total heat generated will then be given by

$$Q = -j\left( \eta + T \frac{\partial U_0}{\partial T} \right) \qquad (15)$$

And the temperature dependence of open circuit voltage is given by

$$U_0 = 0.0025T + 0.2329 \qquad (16)$$

The rise in temperature affects the reaction rate as the exchange current density is assumed to be related to the temperature as:

$$aj_{0,c}^{ref}(T) = aj_{0,c}^{ref}(353K)\exp\left[-16456\left(\frac{1}{T} - \frac{1}{353.15}\right)\right] \quad (17)$$

Flow is induced in the domain by implementing a pressure difference between the inlet and the outlet. Dirichlet boundary conditions are imposed for concentration at the inlet and fully developed condition at the exit. At the inlet, the concentration of oxygen is assumed to be $C_{O_2,ref}$. For temperature, Dirichlet boundary condition, $T=T_{op}$, is given at the inlet. $T_{op}$ is the temperature at which the fuel cell is assumed to be operating. For temperature, the insulated boundary condition is given by the solid surface of the current collector and the fully developed condition is given at outlet.

The effective oxygen diffusion coefficient for the porous medium is found from the Bruggman correlation:

$$D_{O_2,eff} = \phi^{1.5} D_{O_2} \quad (18)$$

The thermal diffusivity value in the porous media is calculated as:

$$\alpha_{eff} = \left[k_f \phi + k_s(1-\phi)\right]/(\rho c_p)_{eff} \quad (19)$$

$$(\rho c_p)_{eff} = \phi(\rho c_p)_f + (1-\phi)(\rho c_p)_s$$

Here, the subscript $f$ stands for fluid and $s$ stands for the solid phase.

## 5. RESULTS AND DISCUSSION

Simulations are conducted for different operating temperatures and operating voltages for various flow rates in PEMFCs with different values of porosity, permeability and gas diffusivity of the porous gas diffusion layer (GDL). Flow rates are varied by varying the specified pressure at the outlet. Reynolds numbers (Re) for each flow rate is calculated based on the average velocity, given by the ratio of the volume flow rate and the cross-sectional area, and is given in Tab. 1. Study of the effects of operating voltage, operating temperature, the porosity and permeability of the GDL, the gas diffusion rate in the porous layer and flow rate on the performance and operation characteristics is conducted. In the plots and discussions, the concentrations and temperature are non-dimensionalized with their respective reference values. It is assumed that the water produced will remain in the vapor state facilitating single phase assumption throughout the domain. The geometry of the electrode is assumed as shown in the schematic in figure 4.

The contours of flow direction velocity and the oxygen concentration are shown in figure 5 for Re=Re1, $\phi$=0.4 and $T_{op}$=353 K. Figure 5(a) shows that the flow velocity in the porous layer is very small compared to that in the flow channel. This suggests that the transport in the porous GDL will be diffusion dominant. Figures 5(b), 5(c) and 5(d) show the steady state oxygen concentration distribution for three operating voltages. With decreasing operating voltage, the oxygen concentration near the reaction surface is seen to decrease. This is attributed to the increase in the reaction rate at the catalyst later resulting in the increased output current density.

For operating voltages above 0.6 V, at this inlet flow rate value, the oxygen concentration is seen to be more or less uniform with adequate oxygen concentration near the catalyst layer. This is due to the lower current density at the high operating voltages while the higher flow rate ensures better transport. However, for lower operating voltage values, below 0.55 V, the oxygen concentration near the catalyst layer starts to get depleted. This means that, under these operating conditions, the rate of oxygen transport is not adequate to balance the rate of its consumption due to electrochemical reactions.

The increased reaction rate with decreasing operating voltage also results in the increase in the heat generation rate. This, in turn, results in higher temperature near the catalyst layer. Figure 6(a), 6(b) and 6(c) shows that the temperature increases near the catalyst layer with decreasing operating voltage. The temperature distribution shows that in all the cases, the location of highest temperature is near the flow channel outlet, close to the catalyst layer. Another interesting observation from the temperature distribution contours is that the temperature transport rate is slower in the porous layer when compared to the flow channel. The slower transport in the porous layer is the result of the difference in the thermal diffusion coefficient and negligible convection transport when compared to the flow channel. This results in larger thermal gradients in the porous GDL making it prone to thermal stresses, especially near the outlet region.

The variation of oxygen concentration and temperature in the catalyst layer along the flow direction is plotted in figure 7. The decrease in the oxygen concentration and the increase in the local temperature values is seen from the figure. This confirms that the rate of reaction increases with lowering operating voltage. It is interesting to note that the concentration of oxygen has reached near 0.2 times the inlet value for 0.5 V. Further decrease in the operating voltage results in the increase in the current density and the reduction in oxygen concentration. This lack of oxygen available for reaction at the catalyst layer limits the maximum value of current density achievable. Hence the mass transport limitation restricts the maximum value of current density achievable.

The effect of this transport limitation can be observed from the distribution of current density along the flow direction, shown in figure 8. For higher operating voltages when the current density magnitude is less, its distribution along the length is more or less uniform ($V_{op}$=0.6V & 0.65 V). However, for lower operating voltages, the current density decreases towards the outlet. This is due to the higher consumption of oxygen near the inlet side leading to a lack of its availability to the reaction sites near the outlet.

The variation of output current density and the heat generation rate with operating voltage can be analyzed from the performance curve plotted in figure 9. It is seen that the power output of the electrode is negligible for operating voltage above 0.7 V for the electrode kinetics selected. For lower operating voltages, the current density increases exponentially. However, the maximum value of the power output is limited by the mass transport limitation as for the operating voltage decreases as discussed earlier. The effect of various operating and material parameters on this transport limitation and hence, on the performance, is analyzed in the coming paragraphs.

The first parameter considered is the operating temperature. Simulations are conducted for three operating temperatures namely, 323K, 353K, and 373K. The results show that the output current density of the cell increases with increasing temperature. Figure 10 shows the oxygen concentration and temperature distribution along the catalyst layer for the three operating temperatures. The figure shows that the oxygen concentration is lower for higher operating temperatures. This is due to the increased current density at a higher temperature as seen from figure 11. The performance improvement with temperature can be viewed from figure 12. However, the figure also shows that the heat generation rate also increases with higher operating temperatures. The maximum temperature in the electrode also rises during operation at higher temperatures.

Moreover, figure 11 also shows that the increase in current density between 373K and 353K is less when compared to that between 353K and 323K. The reason behind this is that as the current density increases, the mass transfer limitation becomes significant reducing the reaction rate near the outlet side of the electrode. This is reflected in the performance curve in figure 12 also. The rate of increase of output current density lowers for higher temperature case as the operating voltage changes from 0.55V to 0.5V. Hence the mass transport limitation should be addressed for improving the maximum rate of power output from the cell.

To understand the effect of inlet flow rate on the mass transport limitation, simulations are conducted for three Reynolds numbers, as given in Table 1. The three Reynolds numbers are named $Re_1$, $Re_2$ and $Re_3$, such that $Re_1 > Re_2 > Re_3$. The oxygen concentration distribution in the computational domain is shown in figure 13 for the three Re. The figure shows that the inlet flow rate has a strong influence on the oxygen transport. The contours show that for the case with the lowest flow rate the concentration of oxygen at the catalyst layer near the outlet has almost gone down to 0.05 times the inlet concentration. This means that the reaction rate in this region will also be decreased due to lack of reactant concentration. This can be closely observed from figures 15 and 16. The decrease in oxygen concentration with length for the three flow rates can be seen from figure 15. For $Re_3$, the concentration profile dips sharply with length and the current density profile, in figure 16, also shows the same trend as the concentration profile.

The effect of this transport limitation on performance is seen from figure 17. For $Re_2$, the output current density decreases from that of $Re_1$, for operating voltage 0.5. This shows that the effect of mass transport limitation becomes significant at this operating voltage at this flow rate. However, for $Re_3$, the performance degradation is profound even at higher operating voltages. It is seen that the current density value decreases as the operating voltage changes from 0.55 to 0.5 at this flow rate. The effect of decrease of current density on output power density is plotted in figure 18. With the operation of the cell at 0.5V, the output power density decreases as current density and the operating voltage decreases for $Re_1$. Even for $Re_2$, the power density increase is minimal as the operating voltage drops and the increase in current density is also not substantial.

The reduced flow rate has an adverse effect not only on performance but also on the heat transport. It is seen from figure 17 that with decreasing flow rate, the heat generation rate is also decreased due to lower reaction rates. However, figures 14 and 15 show that the maximum temperature in the GDL increased with lower flow rates. This shows that the heat removal from the system gets seriously affected due to reduced flow rate.

These results depict that increased flow rate aids better performance, heat removal and iso thermalisation. However, it can be noticed from the contours in figure 13 that a large amount of oxygen is going out of the system unused, for higher flow rates. This means that the fuel utilization is seriously affected. Hence optimization of the flow rates and channel geometry is necessary to obtain maximum performance minimizing fuel wastage.

To study the effect of porous medium properties on the mass transport for a fixed flow rate, simulations are conducted for various values of permeability, porosity and effective oxygen diffusion coefficients in the porous GDL. Although these parameters are not independent, each of them is varied without changing others, so that their effect can be studied independently. The distribution of oxygen concentration and temperature in the catalyst layer along the flow direction for three values of permeability is shown in figure 19. The corresponding Darcy numbers for the three permeabilities are calculated based on the thickness of the GDL and given in Table 2. The plots show that the distribution of oxygen concentration is not affected significantly by a change in permeability. The same trend is observed in the current density profile shown in figure 20. The dependence of current density on permeability is also observed to be weak. However, even with the minimal improvement in performance, as seen in figure 21, the temperature distribution is affected significantly (Figures 19 and 22). Although the heat generation rate is same for different permeability values, the maximum temperature value in the GDL increased with permeability. The temperature near the inlet decreased and the temperature near the outlet increased, resulting in larger thermal gradients in the porous layer. This will, in turn, lead to larger thermal stresses and chances of hot spot formation. This shows that increasing the permeability of porous GDL adversely affects the life of an electrode increasing the thermal stresses in it while giving no significant improvement in performance, at least in the range of values simulated in the present work.

Figure 23 shows the distribution of oxygen concentration and temperature in the catalyst layer along the flow direction for three different porosity values. It is seen that there is no significant effect of porosity on concentration and temperature values in the catalyst layer. The local current density values and the performance curves also show similar behavior in which the values are almost completely unaffected by the change in porosity. However, these results do not take into account the change in effective transport coefficients due to change in porosity. The effect of oxygen diffusion coefficients is studied separately from simulations where the porosity and permeability values are kept constant.

Figure 24 shows the variation in oxygen concentration distribution as the diffusion coefficient value in the porous medium changes. It is seen that as the diffusion coefficient decreases, the amount of oxygen transport into the porous GDL is reduced. The variation of oxygen concentration in the direction perpendicular to the flow, at $x=L/2$, is plotted in figure 25 to understand this. The sudden change in slope of the curve occurs at the porous clear interface where porous media is present (in $y<11$ lattice units(l.u.)). The figure shows that, as the diffusion coefficient value decreases, the oxygen concentration gradient in the Y direction increases sharply in the porous layer. This shows that the diffusion rate of oxygen from the flow channel (y>11 l.u.) to the GDL (y<11 l.u.) faces higher resistance. This also leads to lower oxygen concentration at the catalyst layer for reaction. The magnitude of the decrease in reaction rate is seen from the current density distribution along the catalyst layer shown in figure 26. The figure shows that there is a substantial increase in current density throughout the length of the catalyst layer with an increase in diffusivity in porous media. The same can be confirmed from figure 27 where the concentration value along the catalyst layer is plotted. Higher diffusivity confirms the higher concentration of oxygen at the catalyst layer along its entire length. The figure also shows that the temperature value also increases with diffusivity value. This is attributed to the higher heat generation associated with the higher current density.

Figure 25 also shows that the concentration of oxygen in the channel is also higher for low porous media diffusion coefficient values. This is due to the resistance of oxygen diffusion into the porous media from the flow channel. This will, in turn, result in higher reactant flow out through the channel exit, adversely affecting the fuel utilization.

The increase in performance with an increase in diffusivity is seen to be substantial from the performance curve in figure 28. The improvement is especially prominent in the lower operating voltages as in those operating conditions the mass transport limitations are more profound. The importance of higher diffusion rates can be better appreciated from the power density values plotted in figure 29. The output power density increase has significant improvement due to the increased diffusion rate. The improvement is primarily in the lower operating voltage conditions. This means that the transport limitation limiting the maximum power output can be overcome to an extent by the use of GDL materials having better diffusivity values for the reactants.

## 6. CONCLUSIONS

Fluid flow, heat and mass transfer through the cathode of a fuel cell is simulated using MRT Lattice Boltzmann method and the performance of the electrode and the transport in the cathode are studied. The results showed that the maximum achievable output power density is limited by the mass transport rate of reactant. Attention is focused on the factors affecting this reactant transport across the GDL to the reaction sites. The work includes performance analysis for different operating voltages, operating temperatures, and flow rates for varying permeability, porosity and oxygen diffusion coefficient values in the porous media.

The results suggest that the performance of the electrode can be enhanced by operating the cell at a higher temperature. However, the maximum output power density is still limited by mass transport of the reactant to the reaction sites. Moreover, operating the cell at higher temperature increases the thermal stresses as the temperature gradients in the GDL also increases.

The porous media properties, namely the permeability and porosity do not exhibit a significant impact on the performance of the cell. On the contrary, higher permeability increased the thermal gradients and chances for hot spot formation in the porous layer.

The two major factors that are seen to improve the performance of the cell are the flow rate and the effective porous media species diffusion coefficient. Higher flow rates facilitate better species transport across the porous layer and hence results in increased power output, especially while operating at higher current densities. It, however, might result in fuel wastage through the flow outlet, adversely affecting the fuel utilization. Optimized flow rate values and channel geometries are required to maximize performance without fuel wastage.

Improved oxygen diffusion coefficient values in the porous layer showed attractive performance enhancement as the oxygen availability at the reaction sites improved substantially. Higher oxygen diffusion coefficients also resulted in better fuel utilization as more oxygen could diffuse into the porous GDL. This implies that porous material with improved transport properties is a major factor in the performance enhancement of fuel cells.

The present study is conducted with the assumption that all the components in the electrode are in a single phase. The transport and hence the impact of each of the parameters studied in the present work can be different when the transport is in multiple phases. Available literature on the impact of multiphase transport in electrodes on the cell performance is still inadequate and should be the direction in which research should be focused in the near future.

**ACKNOWLEDGMENT**

The computation work has been carried out on the computers provided by IITK (www.iitk.ac.in/cc). Data analysis and article preparation has been carried out using the resources available at IITK. This support is deeply appreciated.**NOMENCLATURE**

| | |
|---|---|
| $aj_{0,c}^{ref}$ | Reference exchange current density × area of cathode, $1\times10^{-4}$ A/cm$^3$ |
| $C_{O_2,ref}$ | Reference oxygen concentration at cathode, $17.808\times10^{-6}$ mol/cm$^2$ |
| $\alpha_c$ | Cathodic transfer coefficient, 1.0 |
| F | Faraday constant, 96487 C/mol |
| R | Universal gas constant, 8.314 J/mol K |
| $U_0$ | Thermodynamic equilibrium potential, V |
| $V_{op}$ | Cell operating potential, V |
| $D_{O_2}$ | Oxygen diffusivity, $5.2197\times10^{-2}$ cm$^2$/s |
| $k_f$ | Fluid thermal conductivity, 0.026 W/m-K |
| $k_s$ | Solid thermal conductivity, 1.5 W/m-K |
| $T_{ref}$ | Reference temperature, 353 K |
| $\phi$ | Porosity, 0.4 |
| | |

**REFERENCES**

1. Vasiliev, L.L. and L.L. Vasiliev Jr, *Heat pipes to increase the efficiency of fuel cells.* International Journal of Low-Carbon Technologies, 2009. **4**(2): p. 96-103.
2. Springer, T.E., T. Zawodzinski, and S. Gottesfeld, *Polymer electrolyte fuel cell model.* Journal of the Electrochemical Society, 1991. **138**(8): p. 2334-2342.
3. Bernardi, D.M. and M.W. Verbrugge, *A mathematical model of the solid-polymer-electrolyte fuel cell.* Journal of the Electrochemical Society, 1992. **139**(9): p. 2477-2491.
26

4. Nguyen, T.V. and R.E. White, *A water and heat management model for Proton-Exchange-Membrane fuel cells.* Journal of the Electrochemical Society, 1993. **140**(8): p. 2178-2186.
5. Fuller, T.F. and J. Newman, *Water and thermal management in solid-polymer-electrolyte fuel cells.* Journal of the Electrochemical Society, 1993. **140**(5): p. 1218-1225.
6. Gurau, V., H. Liu, and S. Kakac, *Two-dimensional model for proton exchange membrane fuel cells.* AIChE Journal, 1998. **44**(11): p. 2410-2422.
7. Dutta, S.S., S, *Numerical prediction of temperature distribution in PEM fuel cells.* Numerical Heat Transfer: Part A: Applications, 2000. **38**(2): p. 111-128.
8. Ju, H., et al., *Nonisothermal modeling of polymer electrolyte fuel cells I. Experimental validation.* Journal of the Electrochemical Society, 2005. **152**(8): p. A1645-A1653.
9. Birgersson, E., M. Noponen, and M. Vynnycky, *Analysis of a two-phase non-isothermal model for a PEFC.* Journal of the Electrochemical Society, 2005. **152**(5): p. A1021-A1034.
10. Sui, P., S. Kumar, and N. Djilali, *Advanced computational tools for PEM fuel cell design: Part 1. Development and base case simulations.* Journal of Power Sources, 2008. **180**(1): p. 410-422.
11. Sui, P., S. Kumar, and N. Djilali, *Advanced computational tools for PEM fuel cell design: Part 2. Detailed experimental validation and parametric study.* Journal of Power Sources, 2008. **180**(1): p. 423-432.
12. Cao, T.-F., et al., *Numerical investigation of the coupled water and thermal management in PEM fuel cell.* Applied energy, 2013. **112**: p. 1115-1125.
13. Xing, L., et al., *A two-phase flow and non-isothermal agglomerate model for a proton exchange membrane (PEM) fuel cell.* Energy, 2014. **73**: p. 618-634.
14. Wang, C.-Y., *Fundamental models for fuel cell engineering.* Chemical reviews, 2004. **104**(10): p. 4727-4766.
15. Bednarek, T. and G. Tsotridis, *Issues associated with modelling of proton exchange membrane fuel cell by computational fluid dynamics.* Journal of Power Sources, 2017. **343**: p. 550-563.
16. Chen, S. and G.D. Doolen, *Lattice Boltzmann method for fluid flows.* Annual review of fluid mechanics, 1998. **30**(1): p. 329-364.
17. Qian, Y.-H., S. Succi, and S. Orszag, *Recent advances in lattice Boltzmann computing.* Annu. Rev. Comput. Phys, 1995. **3**: p. 195-242.
18. Seta, T., E. Takegoshi, and K. Okui, *Lattice Boltzmann simulation of natural convection in porous media.* Mathematics and Computers in Simulation, 2006. **72**(2): p. 195-200.
19. Chai, Z., Z. Guo, and B. Shi, *Lattice Boltzmann simulation of mixed convection in a driven cavity packed with porous medium.* Computational Science–ICCS 2007, 2007: p. 802-809.
20. Wang, L.-P. and B. Afsharpoya, *Modeling fluid flow in fuel cells using the lattice-Boltzmann approach.* Mathematics and Computers in Simulation, 2006. **72**(2): p. 242-248.
21. Molaeimanesh, G., H.S. Googarchin, and A.Q. Moqaddam, *Lattice Boltzmann simulation of proton exchange membrane fuel cells–A review on opportunities*

  *and challenges.* International Journal of Hydrogen Energy, 2016. **41**(47): p. 22221-22245.
22. Park, J. and X. Li, *Multi-phase micro-scale flow simulation in the electrodes of a PEM fuel cell by lattice Boltzmann method.* Journal of Power Sources, 2008. **178**(1): p. 248-257.
23. Mukherjee, P.P., C.-Y. Wang, and Q. Kang, *Mesoscopic modeling of two-phase behavior and flooding phenomena in polymer electrolyte fuel cells.* Electrochimica Acta, 2009. **54**(27): p. 6861-6875.
24. Hao, L. and P. Cheng, *Lattice Boltzmann simulations of water transport in gas diffusion layer of a polymer electrolyte membrane fuel cell.* Journal of Power Sources, 2010. **195**(12): p. 3870-3881.
25. Rama, P., et al., *A numerical study of structural change and anisotropic permeability in compressed carbon cloth polymer electrolyte fuel cell gas diffusion layers.* Fuel Cells, 2011. **11**(2): p. 274-285.
26. Yablecki, J., J. Hinebaugh, and A. Bazylak, *Effect of liquid water presence on PEMFC GDL effective thermal conductivity.* Journal of the Electrochemical Society, 2012. **159**(12): p. F805-F809.
27. García-Salaberri, P.A., et al., *Effective diffusivity in partially-saturated carbon-fiber gas diffusion layers: Effect of through-plane saturation distribution.* International Journal of Heat and Mass Transfer, 2015. **86**: p. 319-333.
28. Van Genuchten, M.T., *A closed-form equation for predicting the hydraulic conductivity of unsaturated soils.* Soil science society of America journal, 1980. **44**(5): p. 892-898.
29. Ostadi, H., et al., *3D reconstruction of a gas diffusion layer and a microporous layer.* Journal of Membrane Science, 2010. **351**(1): p. 69-74.
30. Kim, K.N., et al., *Lattice Boltzmann simulation of liquid water transport in microporous and gas diffusion layers of polymer electrolyte membrane fuel cells.* Journal of Power Sources, 2015. **278**: p. 703-717.
31. Kim, S.H. and H. Pitsch, *Reconstruction and effective transport properties of the catalyst layer in PEM fuel cells.* Journal of the Electrochemical Society, 2009. **156**(6): p. B673-B681.
32. Gao, Y., *Using MRT lattice Boltzmann method to simulate gas flow in simplified catalyst layer for different inlet–outlet pressure ratio.* International Journal of Heat and Mass Transfer, 2015. **88**: p. 122-132.
33. Chen, L., et al., *Lattice Boltzmann pore-scale investigation of coupled physical-electrochemical processes in C/Pt and non-precious metal cathode catalyst layers in proton exchange membrane fuel cells.* Electrochimica Acta, 2015. **158**: p. 175-186.
34. Tabe, Y., et al., *Numerical simulation of liquid water and gas flow in a channel and a simplified gas diffusion layer model of polymer electrolyte membrane fuel cells using the lattice Boltzmann method.* Journal of Power Sources, 2009. **193**(1): p. 24-31.
35. Chen, L., et al., *Pore-scale flow and mass transport in gas diffusion layer of proton exchange membrane fuel cell with interdigitated flow fields.* International Journal of Thermal Sciences, 2012. **51**: p. 132-144.

36. Molaeimanesh, G. and M. Akbari, *Impact of PTFE distribution on the removal of liquid water from a PEMFC electrode by lattice Boltzmann method.* International Journal of Hydrogen Energy, 2014. **39**(16): p. 8401-8409.
37. Molaeimanesh, G. and M.H. Akbari, *Water droplet dynamic behavior during removal from a proton exchange membrane fuel cell gas diffusion layer by Lattice-Boltzmann method.* Korean Journal of Chemical Engineering, 2014. **31**(4): p. 598-610.
38. Molaeimanesh, G. and M. Akbari, *Role of wettability and water droplet size during water removal from a PEMFC GDL by lattice Boltzmann method.* International Journal of Hydrogen Energy, 2016. **41**(33): p. 14872-14884.
39. Hao, L. and P. Cheng, *Lattice Boltzmann simulations of liquid droplet dynamic behavior on a hydrophobic surface of a gas flow channel.* Journal of Power Sources, 2009. **190**(2): p. 435-446.
40. Han, B., J. Yu, and H. Meng, *Lattice Boltzmann simulations of liquid droplets development and interaction in a gas channel of a proton exchange membrane fuel cell.* Journal of Power Sources, 2012. **202**: p. 175-183.
41. Han, B. and H. Meng, *Lattice Boltzmann simulation of liquid water transport in turning regions of serpentine gas channels in proton exchange membrane fuel cells.* Journal of Power Sources, 2012. **217**: p. 268-279.
42. Salah, Y.B., Y. Tabe, and T. Chikahisa, *Gas channel optimisation for PEM fuel cell using the lattice Boltzmann method.* Energy Procedia, 2012. **28**: p. 125-133.
43. Wu, J. and J.-J. Huang, *Dynamic behaviors of liquid droplets on a gas diffusion layer surface: Hybrid lattice Boltzmann investigation.* Journal of Applied Physics, 2015. **118**(4): p. 044902.
44. Ostadi, H., K. Jiang, and P. Prewett, *Micro/nano X-ray tomography reconstruction fine-tuning using scanning electron microscope images.* Micro & Nano Letters, 2008. **3**(4): p. 106-109.
45. Chen, L., et al., *Coupling between finite volume method and lattice Boltzmann method and its application to fluid flow and mass transport in proton exchange membrane fuel cell.* International Journal of Heat and Mass Transfer, 2012. **55**(13): p. 3834-3848.
46. Chen, L., et al., *Multi-scale modeling of proton exchange membrane fuel cell by coupling finite volume method and lattice Boltzmann method.* International Journal of Heat and Mass Transfer, 2013. **63**: p. 268-283.
47. Kamali, M., et al., *A multi-component two-phase lattice Boltzmann method applied to a 1-D Fischer–Tropsch reactor.* Chemical engineering journal, 2012. **207**: p. 587-595.
48. Molaeimanesh, G. and M. Akbari, *A three-dimensional pore-scale model of the cathode electrode in polymer-electrolyte membrane fuel cell by lattice Boltzmann method.* Journal of Power Sources, 2014. **258**: p. 89-97.
49. Molaeimanesh, G. and M. Akbari, *Agglomerate modeling of cathode catalyst layer of a PEM fuel cell by the lattice Boltzmann method.* International Journal of Hydrogen Energy, 2015. **40**(15): p. 5169-5185.
50. Guo, Z. and T. Zhao, *A lattice Boltzmann model for convection heat transfer in porous media.* Numerical Heat Transfer, Part B, 2005. **47**(2): p. 157-177.

51. Pan, C., L.-S. Luo, and C.T. Miller, *An evaluation of lattice Boltzmann schemes for porous medium flow simulation.* Computers & fluids, 2006. **35**(8): p. 898-909.
52. Contrino, D., et al., *Lattice-Boltzmann simulations of the thermally driven 2D square cavity at high Rayleigh numbers.* Journal of Computational Physics, 2014. **275**: p. 257-272.
53. Roy, S. and T. Basak, *Finite element analysis of natural convection flows in a square cavity with non-uniformly heated wall (s).* International Journal of Engineering Science, 2005. **43**(8): p. 668-680.
54. Kang, Q., P.C. Lichtner, and D. Zhang, *Lattice Boltzmann pore-scale model for multicomponent reactive transport in porous media.* Journal of Geophysical Research: Solid Earth, 2006. **111**(B5).
55. Jithin, M., M.K. Das, and A. De, *Lattice Boltzmann Simulation of Lithium Peroxide Formation in Lithium–Oxygen Battery.* Journal of Electrochemical Energy Conversion and Storage, 2016. **13**(3): p. 031003.
56. Jithin, M., et al., *Numerical study of bifurcating flow through sudden expansions: effect of divergence and geometric asymmetry.* International Journal of Advances in Engineering Sciences and Applied Mathematics, 2016. **8**(4): p. 259-273.
57. Jithin, M., et al., *Estimation of Permeability of Porous Material Using Pore Scale LBM Simulations*, in *Fluid Mechanics and Fluid Power–Contemporary Research*2017, Springer. p. 1381-1388.
58. Gu, W. and C. Wang, *Thermal-electrochemical modeling of battery systems.* Journal of the Electrochemical Society, 2000. **147**(8): p. 2910-2922.
59. Um, S., C.Y. Wang, and K. Chen, *Computational fluid dynamics modeling of proton exchange membrane fuel cells.* Journal of the Electrochemical society, 2000. **147**(12): p. 4485-4493.
30

## Table Caption List

Table 1    Reynolds number calculation for three flow rates simulated.

Table 2    Darcy number calculation for three permeabilities of the porous layer simulated

## Figure Captions List

Fig. 1    COMPARISON OF FLOW AND HEAT TRANSPORT RESULTS WITH Roy and Basak [53]. FIGURE SHOWS LOCAL NUSSELT NUMBER VARIATION ALONG THE BOTTOM WALL OF A SQUARE CAVITY WITH HEATED WALLS.

Fig. 2    COMPARISON OF RESULTS OF FLOW AND HEAT TRANSPORT IN POROUS MEDIA WITH Guo and Zhao [50]. FIGURE SHOWS TEMPERATURE DISTRIBUTION AT MID-HEIGHT OF A CAVITY FILLED WITH POROUS MEDIUM WITH HEATED WALLS.

Fig. 3    COMPARISON OF RESULTS FOR SURFACE REACTION WITH Kang, Lichtner [54]. FIGURE SHOWS CONCENTRATION PROFILES OF A SPECIES UNDERGOING SURFACE REACTION AT THE TOP WALL.

Fig. 4    2D SCHEMATIC OF PEMFC. COMPUTATIONAL DOMAIN SHOWN IN DOTTED RECTANGLE (FIGURE NOT TO SCALE).

Fig. 5    (a) CONTOURS OF FLOW DIRECTION VELOCITY FOR Re=$Re_1$. (b), (c) AND (d) ARE CONTOURS OF OXYGEN CONCENTRATION FOR Re=$Re_1$, $\phi$ =0.4, $D_{O_2,eff}$ =1.32×10$^{-2}$cm$^2$/s, k=0.176×10$^{-6}$cm$^2$ AND $T_{op}$=353K.

Fig. 6    CONTOURS OF TEMPERATURE FOR THREE OPERATING VOLTAGES

Fig. 7    CONCENTRATION AND TEMPERATURE PROFILE ALONG CATALYST LAYER FOR THREE OPERATING VOLTAGES. THE

Table 1: REYNOLDS NUMBER CALCULATION FOR THREE FLOW RATES SIMULATED.

|     | Volume Flow Rate (cm$^3$/s) | $U_{avg}$ (cm/s) | Re |
|-----|-----|-----|-----|
| Re$_1$ | 1.11×10$^{-2}$ | 61.908 | 25.36 |
| Re$_2$ | 6.186×10$^{-3}$ | 34.501 | 14.13 |
| Re$_3$ | 1.237×10$^{-3}$ | 6.899 | 2.826 |

Table 2: DARCY NUMBER CALCULATION FOR THREE PERMEABILITIES OF THE POROUS LAYER SIMULATED.

|     | Permeability (cm$^2$) | Da |
|-----|-----|-----|
| K$_1$ | 0.176×10$^{-6}$ | 2.72×10$^{-4}$ |
| K$_2$ | 0.88×10$^{-6}$ | 1.364×10$^{-3}$ |
| K$_3$ | 1.76×10$^{-6}$ | 2.72×10$^{-3}$ |

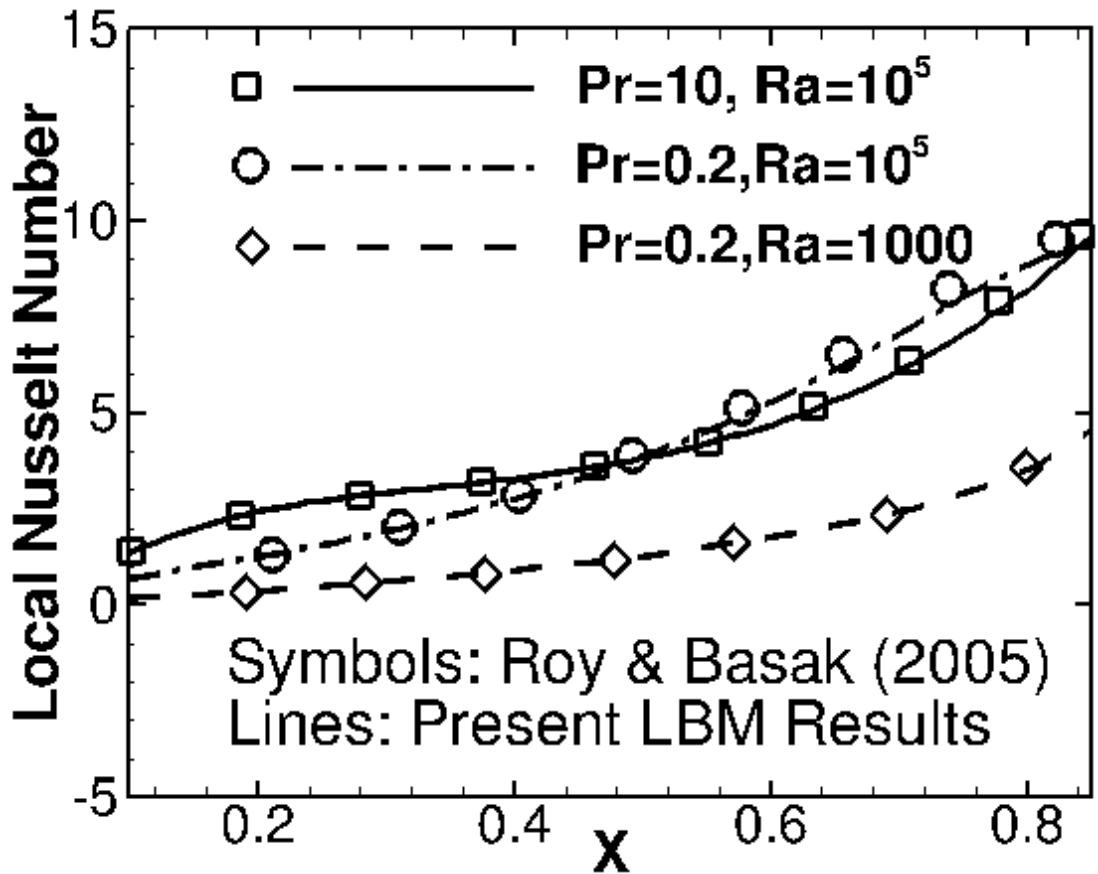

Figure 1. COMPARISON OF FLOW AND HEAT TRANSPORT RESULTS WITH Roy and Basak [53]. FIGURE SHOWS LOCAL NUSSELT NUMBER VARIATION ALONG THE BOTTOM WALL OF A SQUARE CAVITY WITH HEATED WALLS.

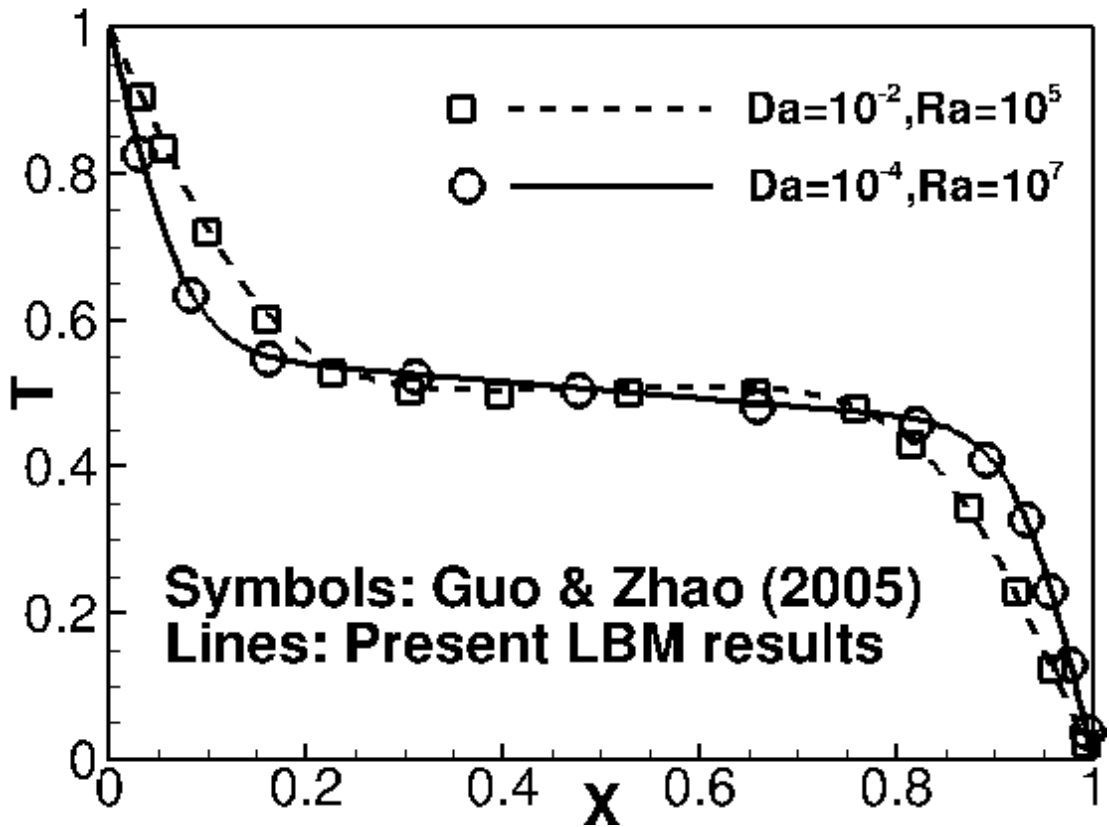

Figure 2. COMPARISON OF RESULTS OF FLOW AND HEAT TRANSPORT IN POROUS MEDIA WITH Guo and Zhao [50Guo and Zhao [50]. FIGURE SHOWS TEMPERATURE DISTRIBUTION AT MID-HEIGHT OF A CAVITY FILLED WITH POROUS MEDIUM WITH HEATED WALLS.

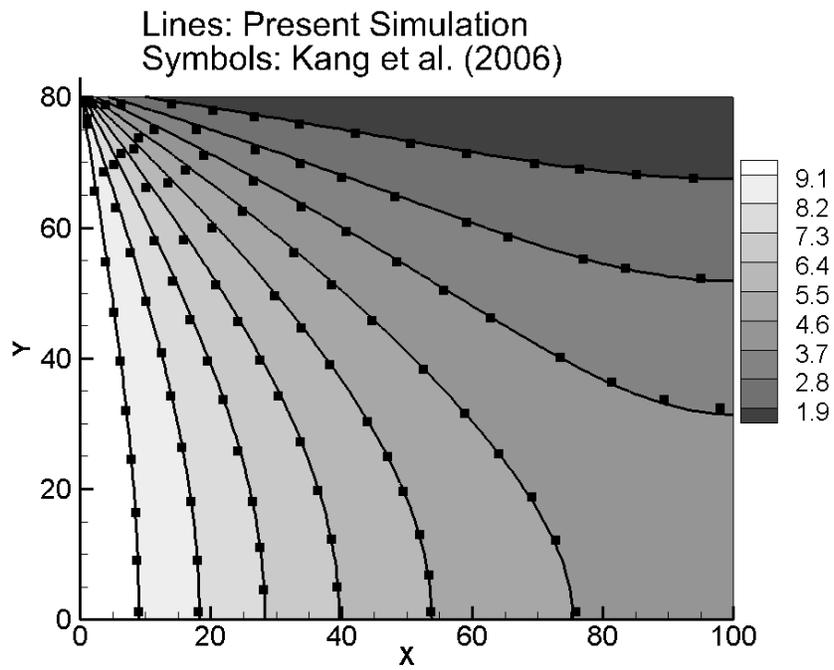

Figure 3. COMPARISON OF RESULTS FOR SURFACE REACTION WITH Kang, Lichtner [54]. FIGURE SHOWS CONCENTRATION PROFILES OF A SPECIES UNDERGOING SURFACE REACTION AT THE TOP WALL.

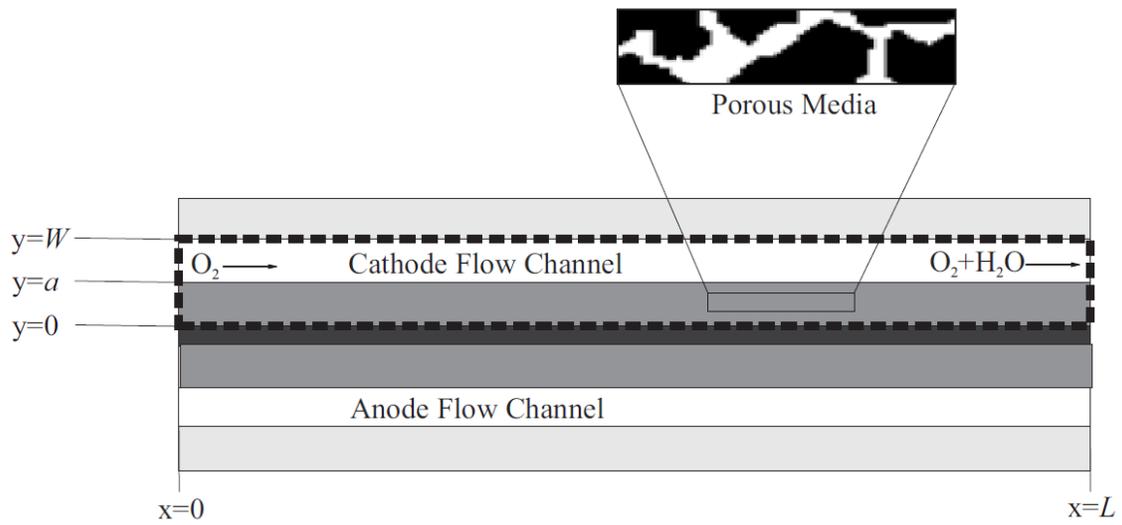

Figure 4. 2D SCHEMATIC OF PEMFC. COMPUTATIONAL DOMAIN SHOWN IN DOTTED RECTANGLE (FIGURE NOT TO SCALE).

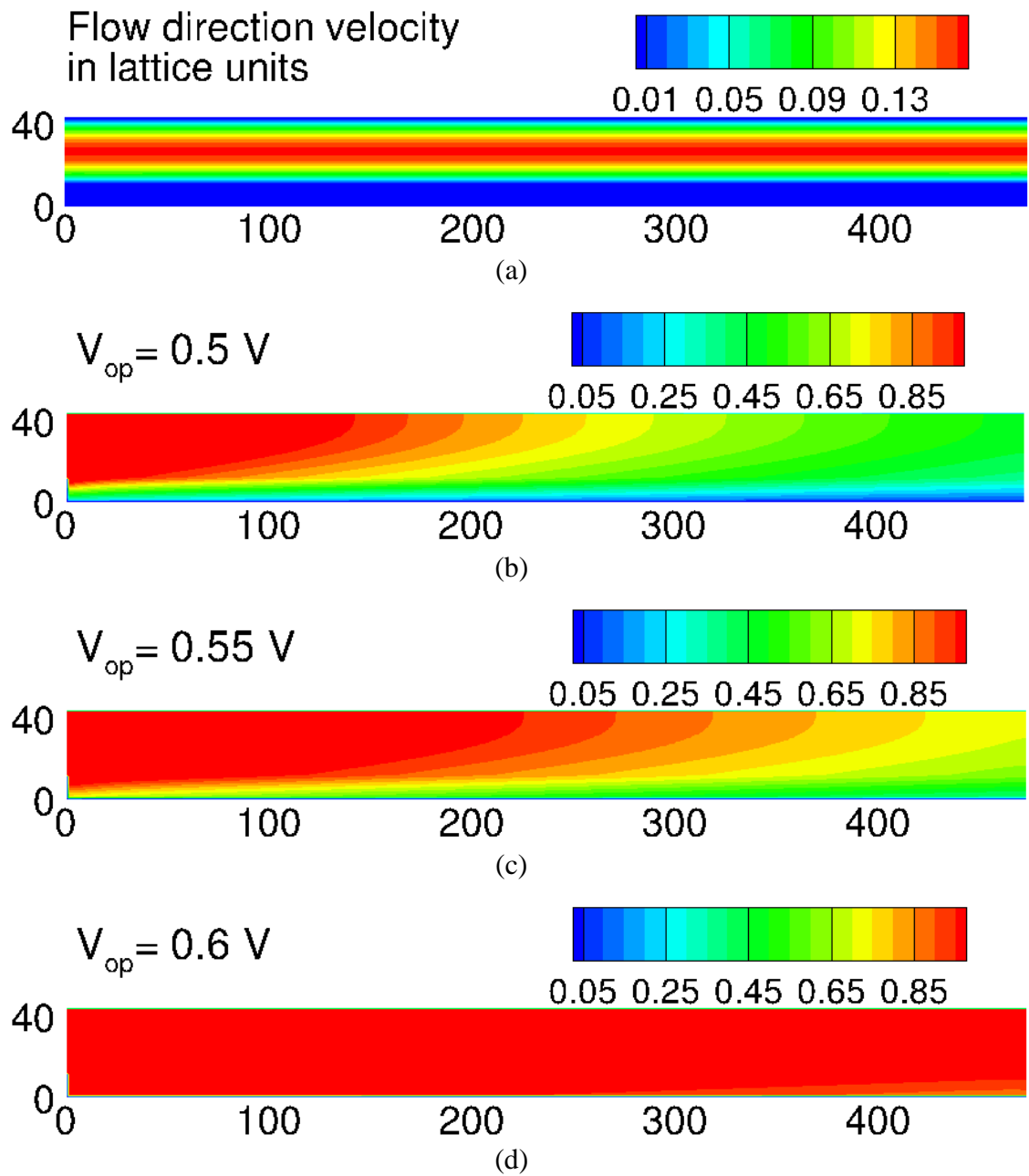

Figure 5. (a) CONTOURS OF FLOW DIRECTION VELOCITY FOR Re=Re$_1$. (b), (c) AND (d) ARE CONTOURS OF OXYGEN CONCENTRATION FOR Re=Re$_1$, $\phi$=0.4, $D_{O_2,eff}$=1.32×10$^{-2}$cm$^2$/s, k=0.176×10$^{-6}$cm$^2$ AND T$_{op}$=353K.

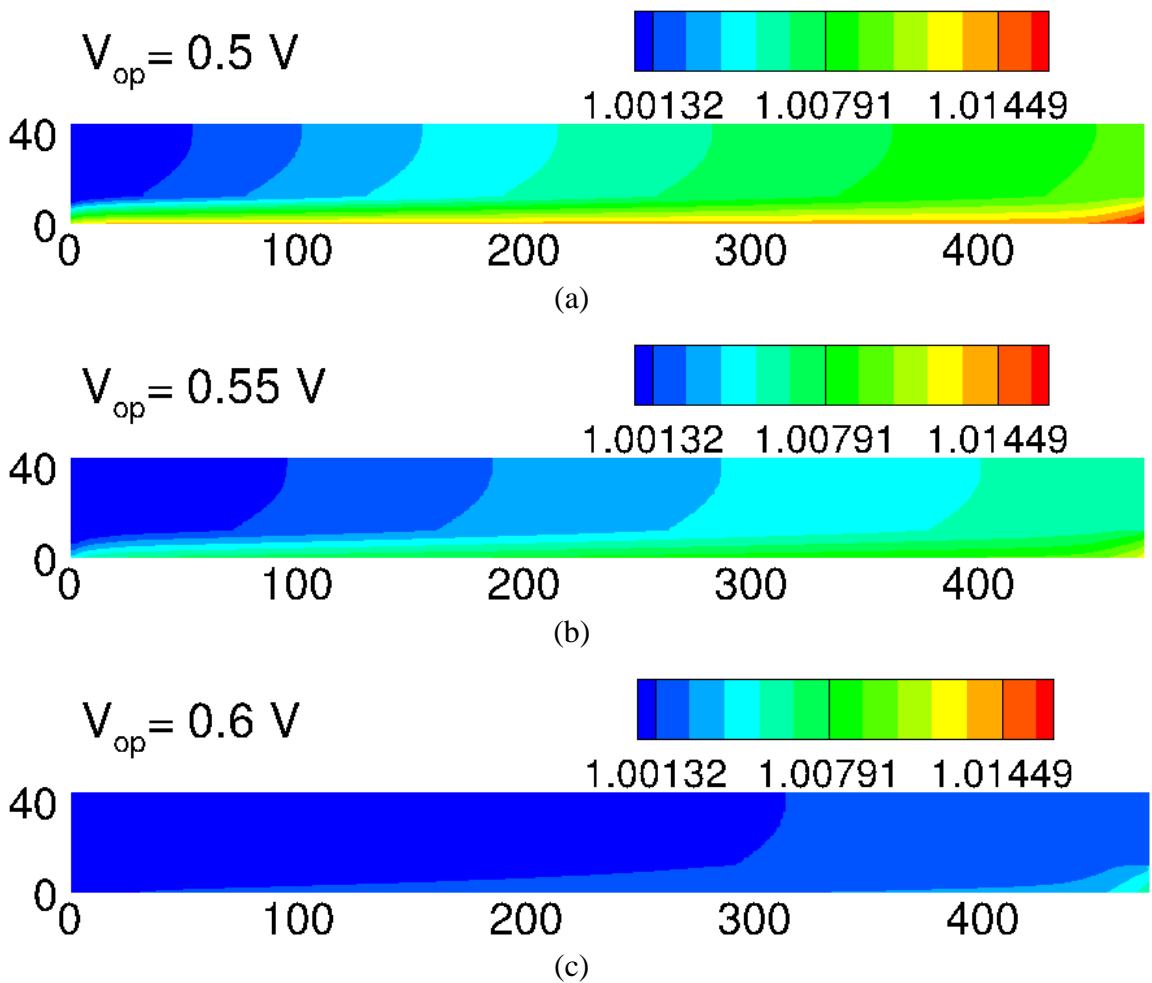

Figure 6: CONTOURS OF TEMPERATURE FOR THREE OPERATING VOLTAGES

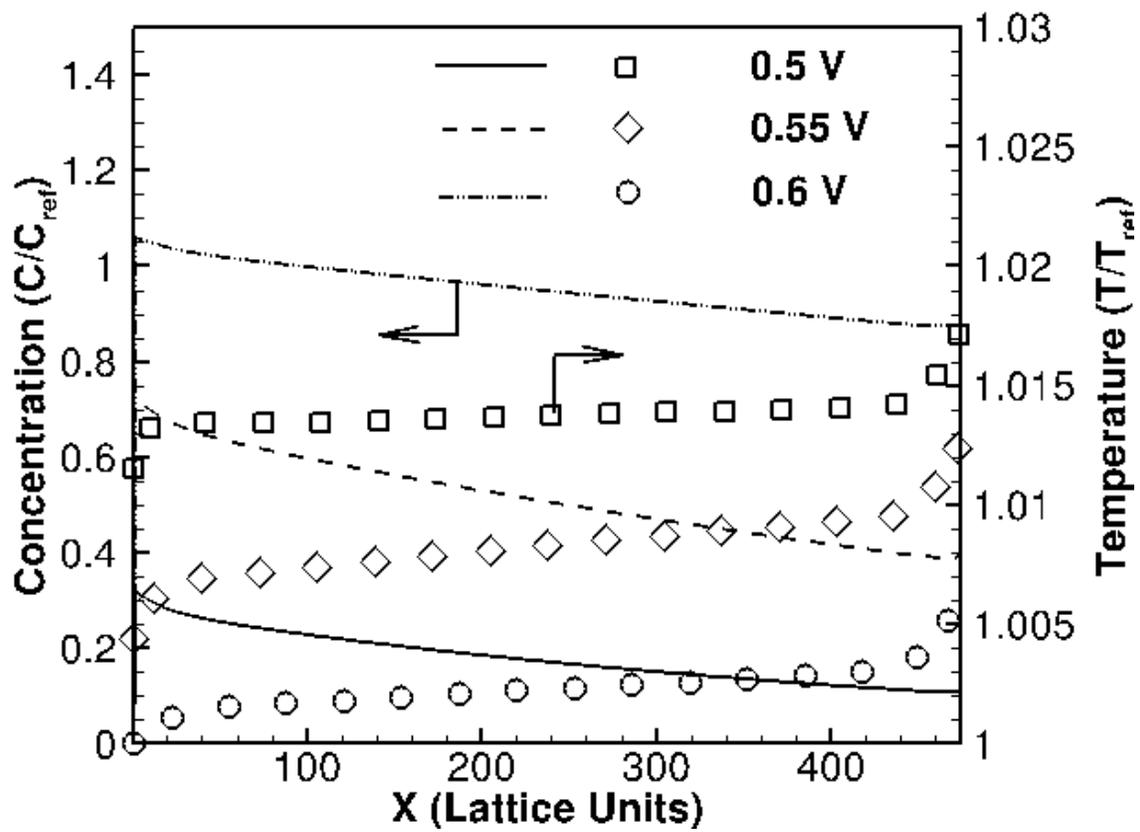

Figure 7. CONCENTRATION AND TEMPERATURE PROFILE ALONG CATALYST LAYER FOR THREE OPERATING VOLTAGES. THE SYMBOLS AND LINES REPRESENT TEMPERATURE AND OXYGEN CONCENTRATIONS RESPECTIVELY.

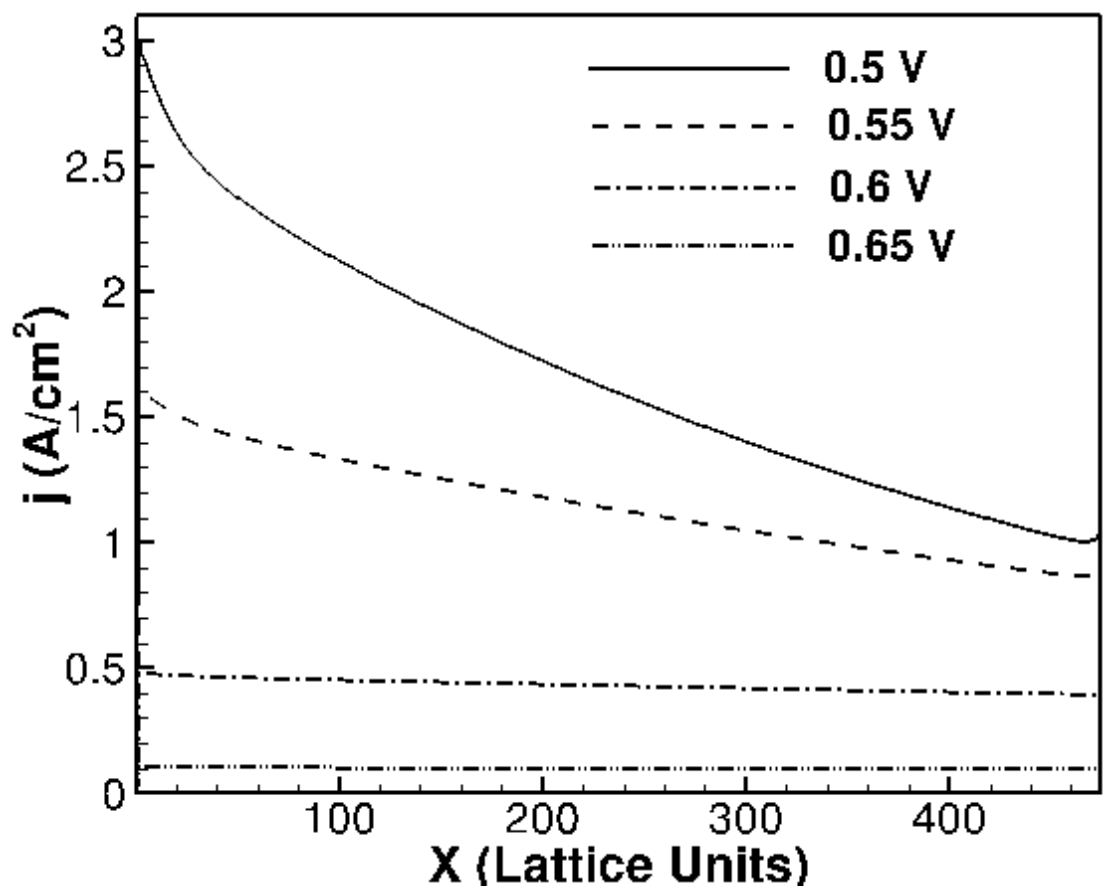

Figure 8. CURRENT DENSITY DISTRIBUTION ALONG CATALYST LAYER FOR DIFFERENT OPERATING VOLTAGES

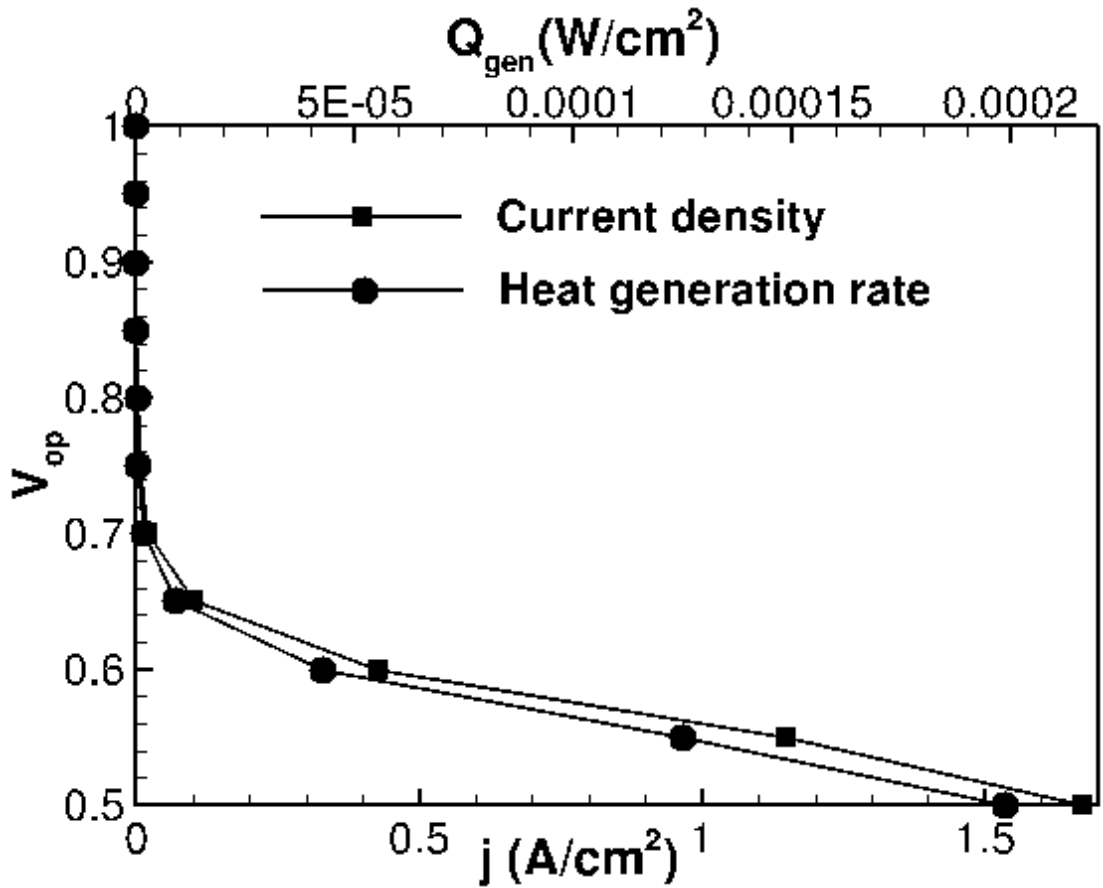

Figure 9. PERFORMANCE CURVE AND HEAT GENERATION ARTE FOR Re=Re$_1$, $\phi$ =0.4 $D_{O_2,eff}$ =1.32×10$^{-2}$cm$^2$/s, k=0.176×10$^{-6}$cm$^2$ AND T$_{op}$=353K.

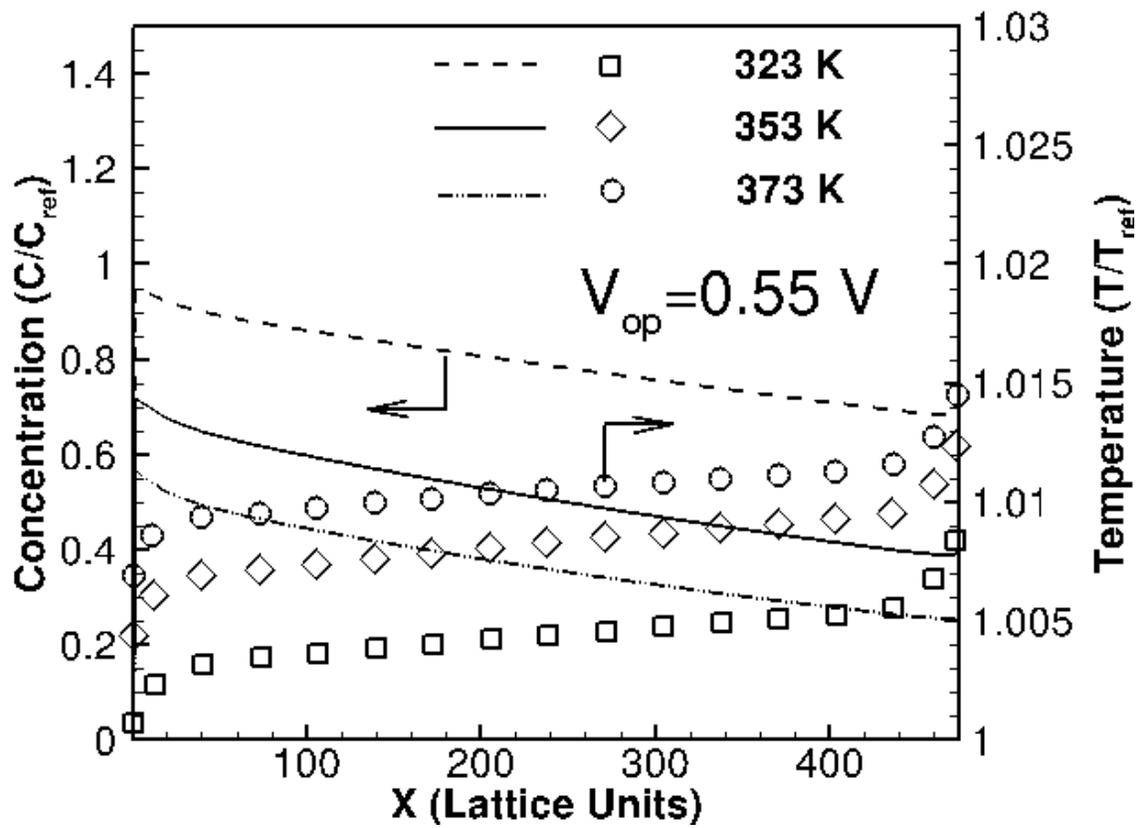

Figure 10. OXYGEN CONCENTRATION AND TEMPERATURE DISTRIBUTION ALONG CATALYST LAYER FOR DIFFERENT TEMPERATURES

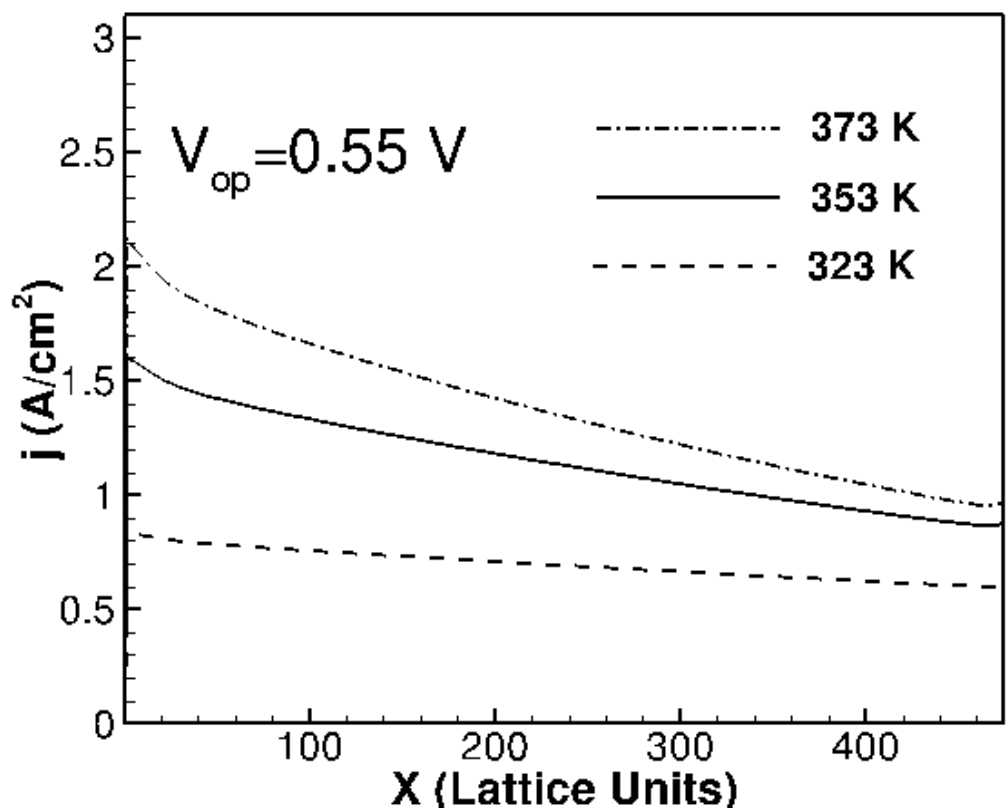

Figure 11. CURRENT DENSITY DISTRIBUTION ALONG CATALYST LAYER FOR DIFFERENT TEMPERATURES

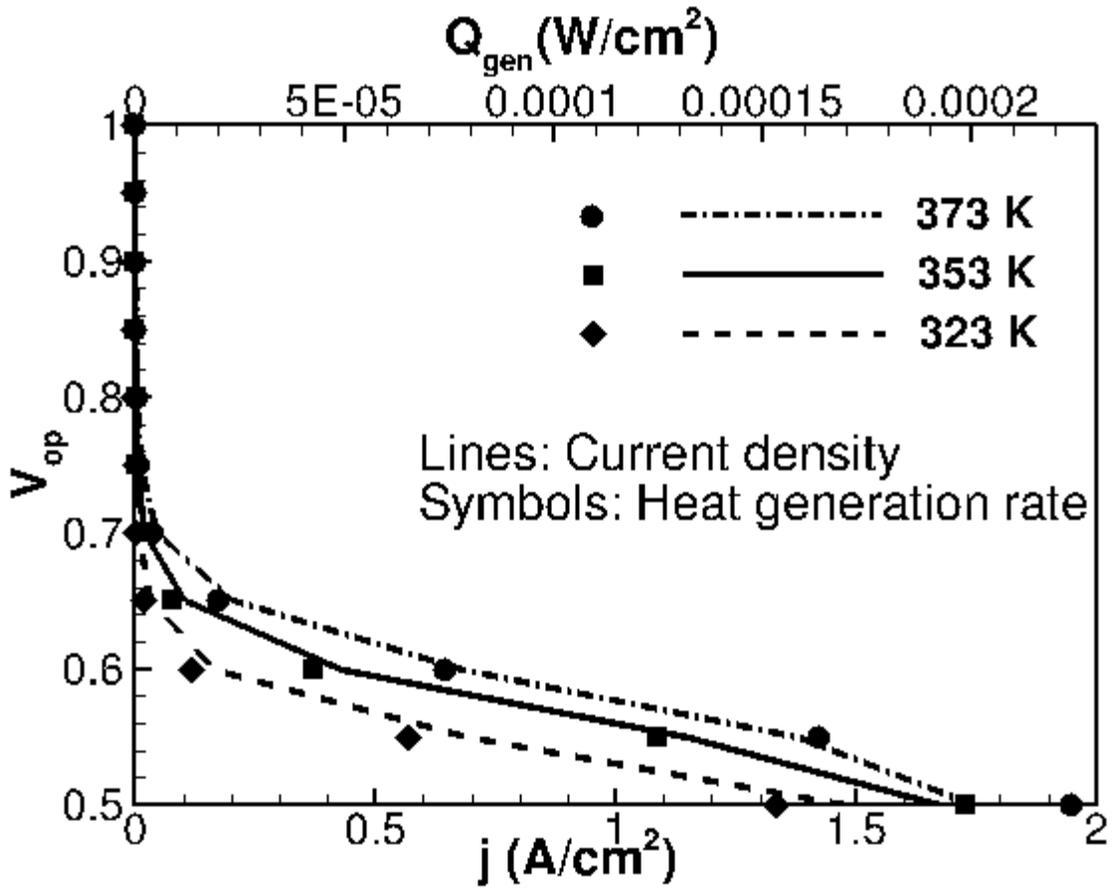

Figure 12. PERFORMANCE AND HEAT GENERATION RATE FOR DIFFERENT OPERATING TEMEPRATURES

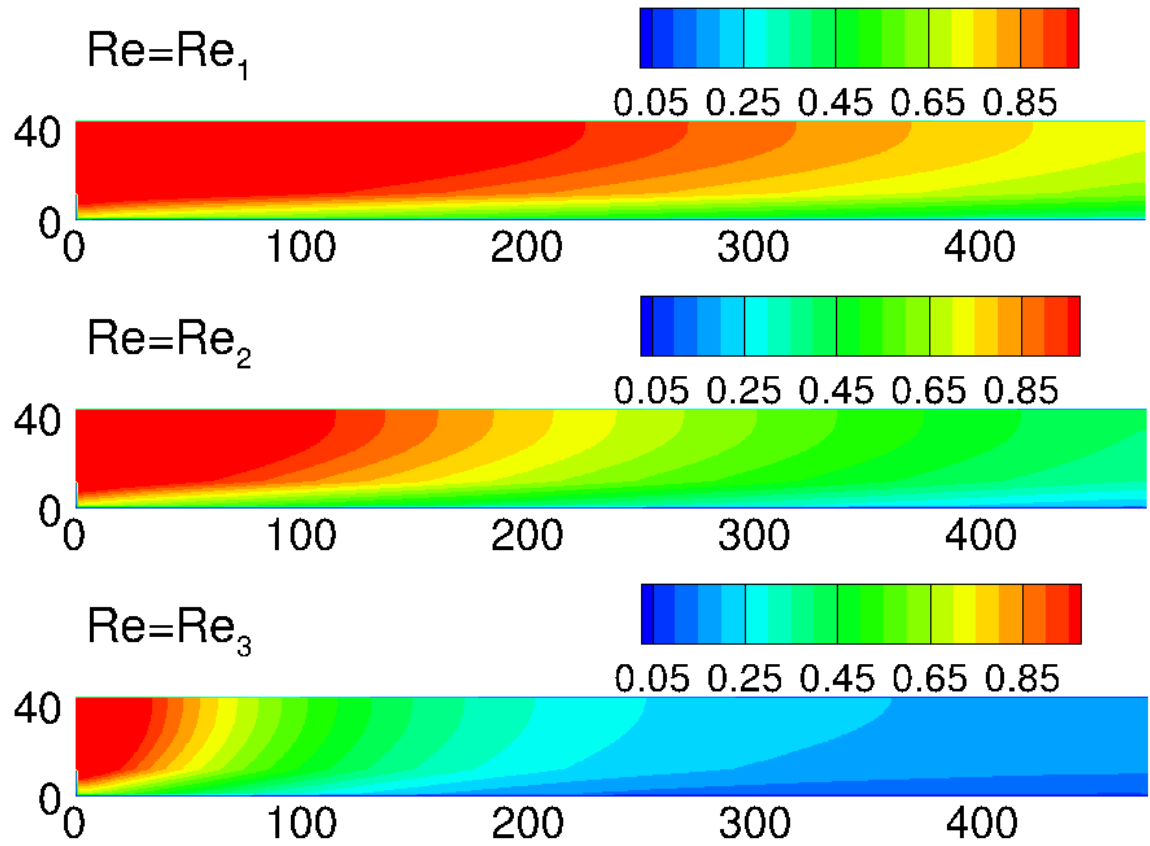

Figure 13. OXYGEN CONCENTRATION CONTOURS FOR DIFFERENT FLOW RATES AT $V_{OP}=0.55V$

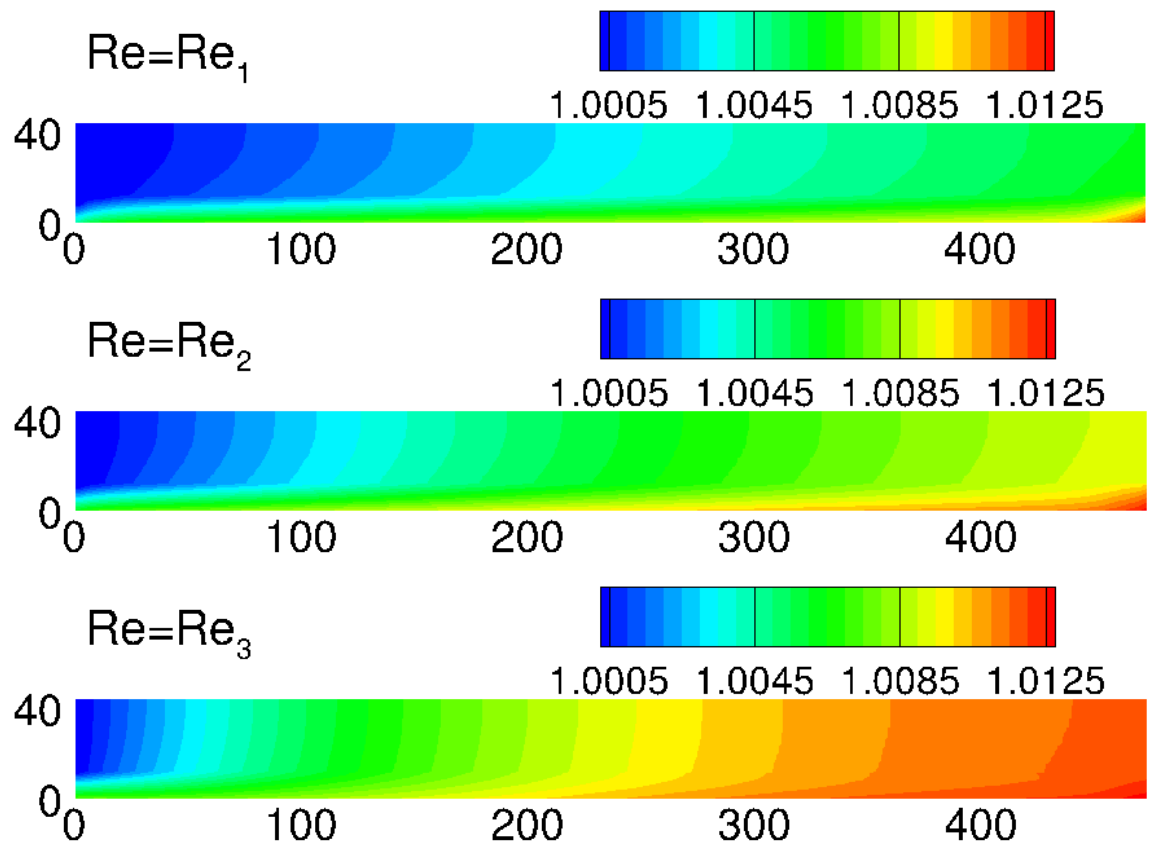

Figure 14. TEMPERATURE CONTOURS AT $V_{op}=0.55V$

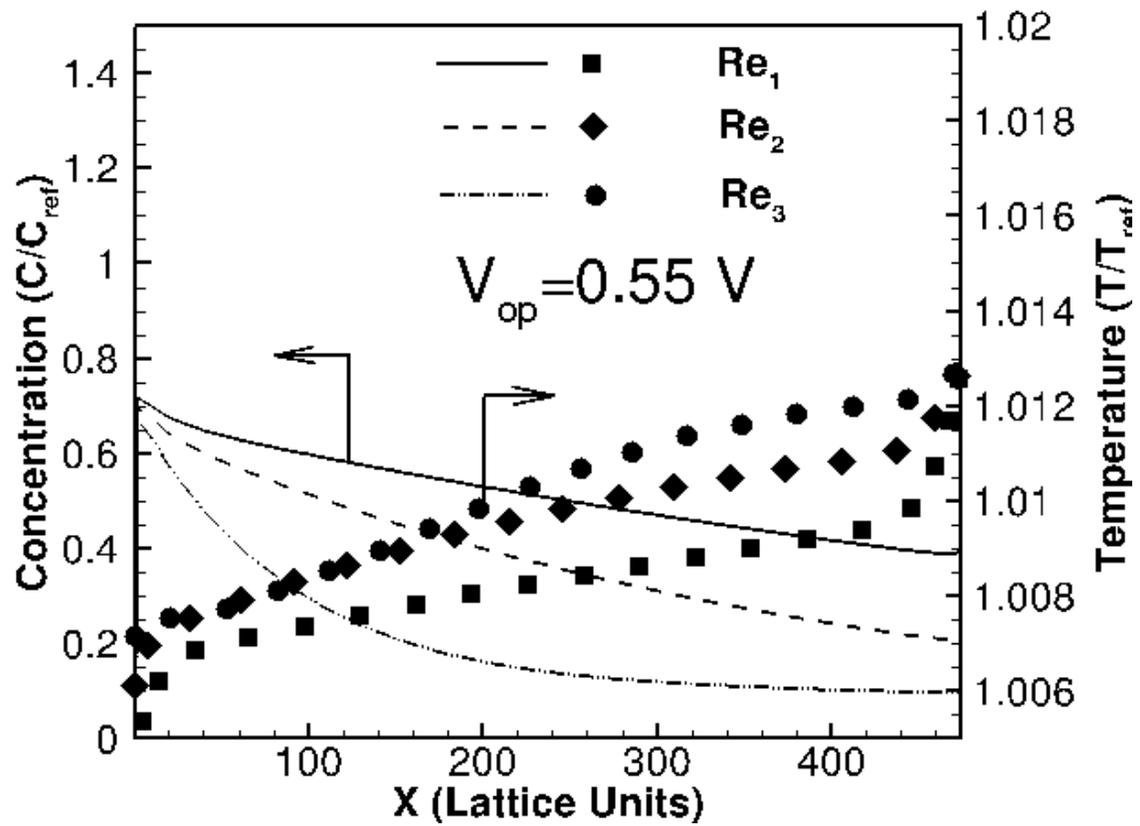

Figure 15. OXYGEN CONCENTRATION AND TEMPERATURE PROFILE ALONG CATALYST LAYER FOR DIFFERENT FLOW RATES AT $V_{OP}=0.55V$

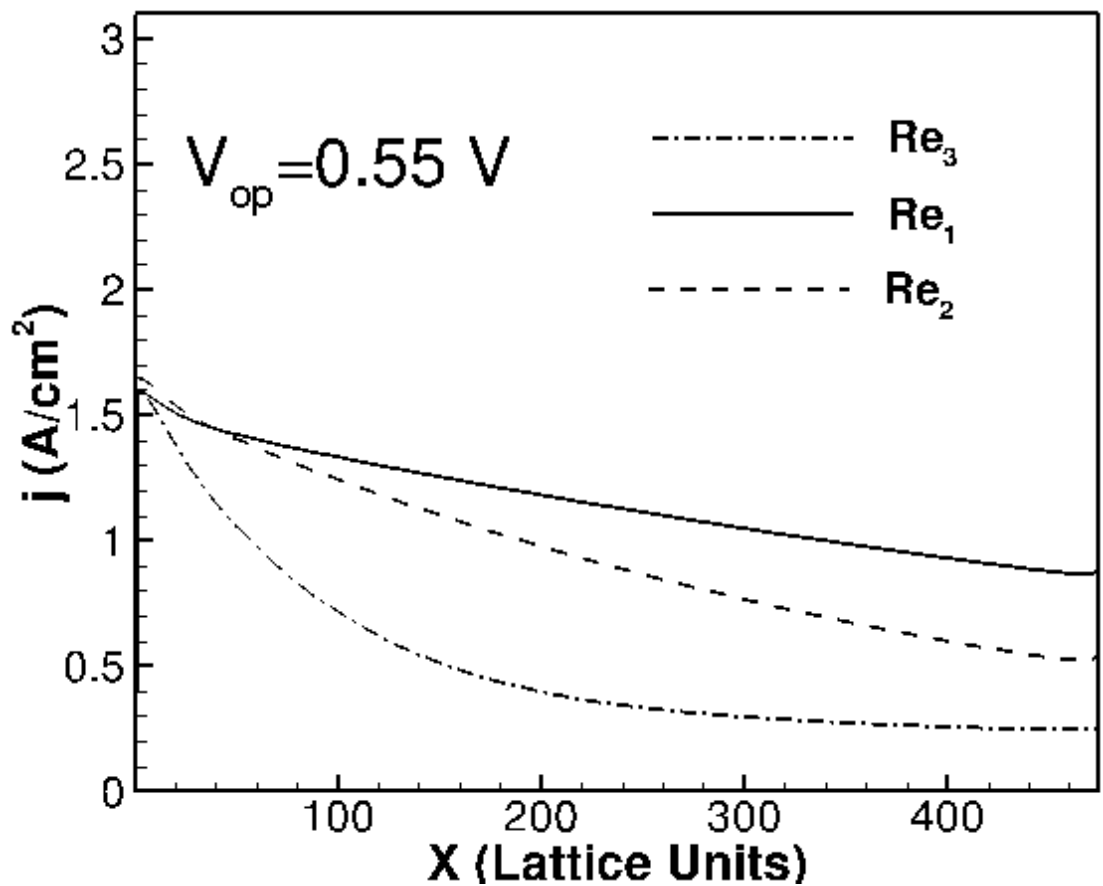

Figure 16: CURRENT DENSITY PROFILE ALONG CATALYST LAYER FOR $V_{op}$=0.55V

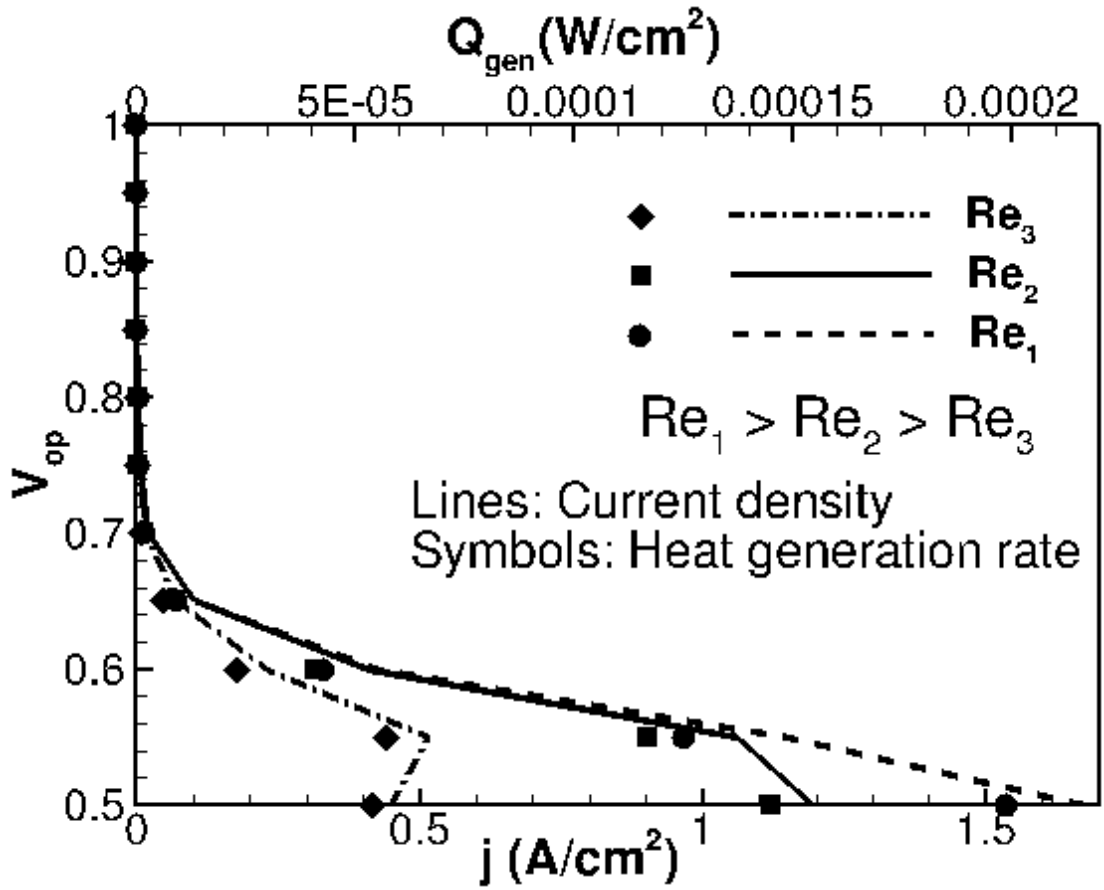

Figure 17: PERFORMANCE AND HEAT GENERATION RATE FOR DFFERENT FLOW RATES

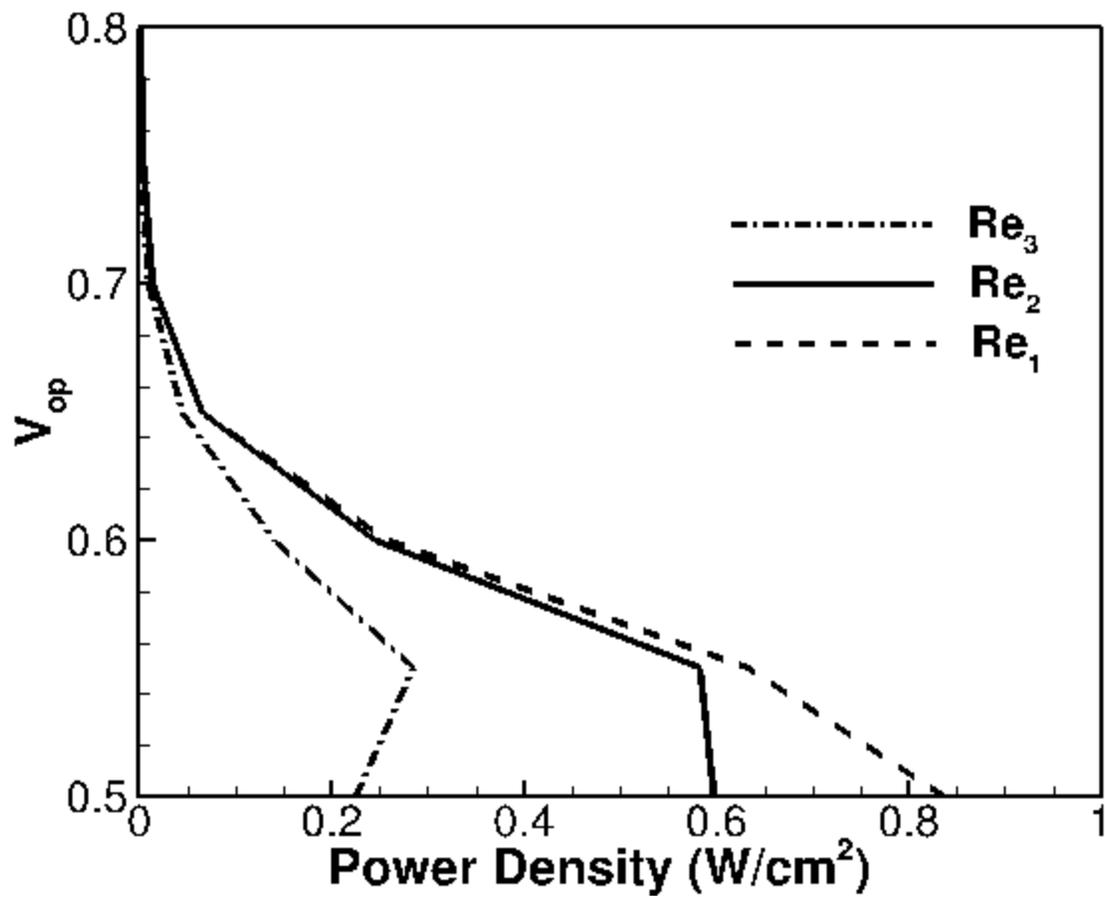

Figure 18: POWER DENSITY VARIATION WITH OPERATING VOLTAGE FOR DIFFERENT FLOW RATES

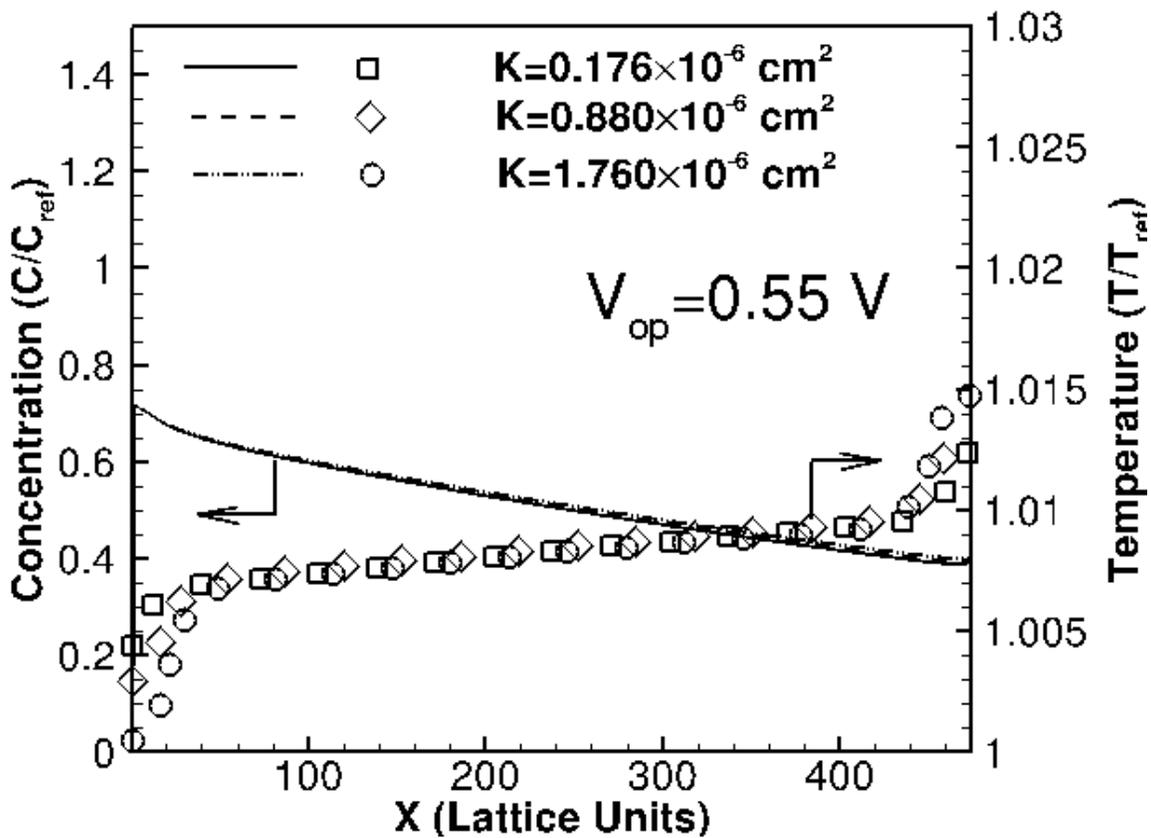

Figure 19: TEMPERATURE AND CONCENTRATION DISTRIBUTION ALONG CATALYST LAYER FOR DIFFERENT PERMEABILITY VALUES

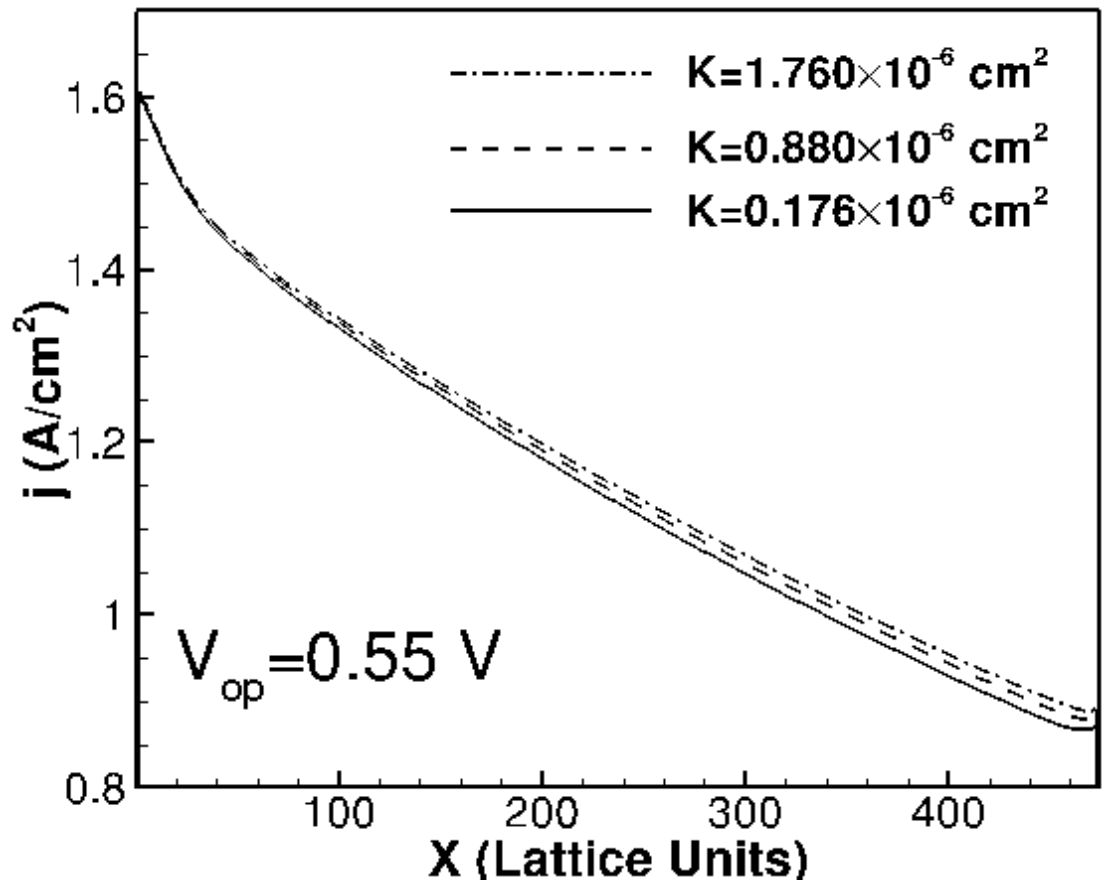

Figure 20: CURRENT DENSITY PROFILE ALONG THE CATALYST LAYER FOR DIFFERENT PERMEABILITIES

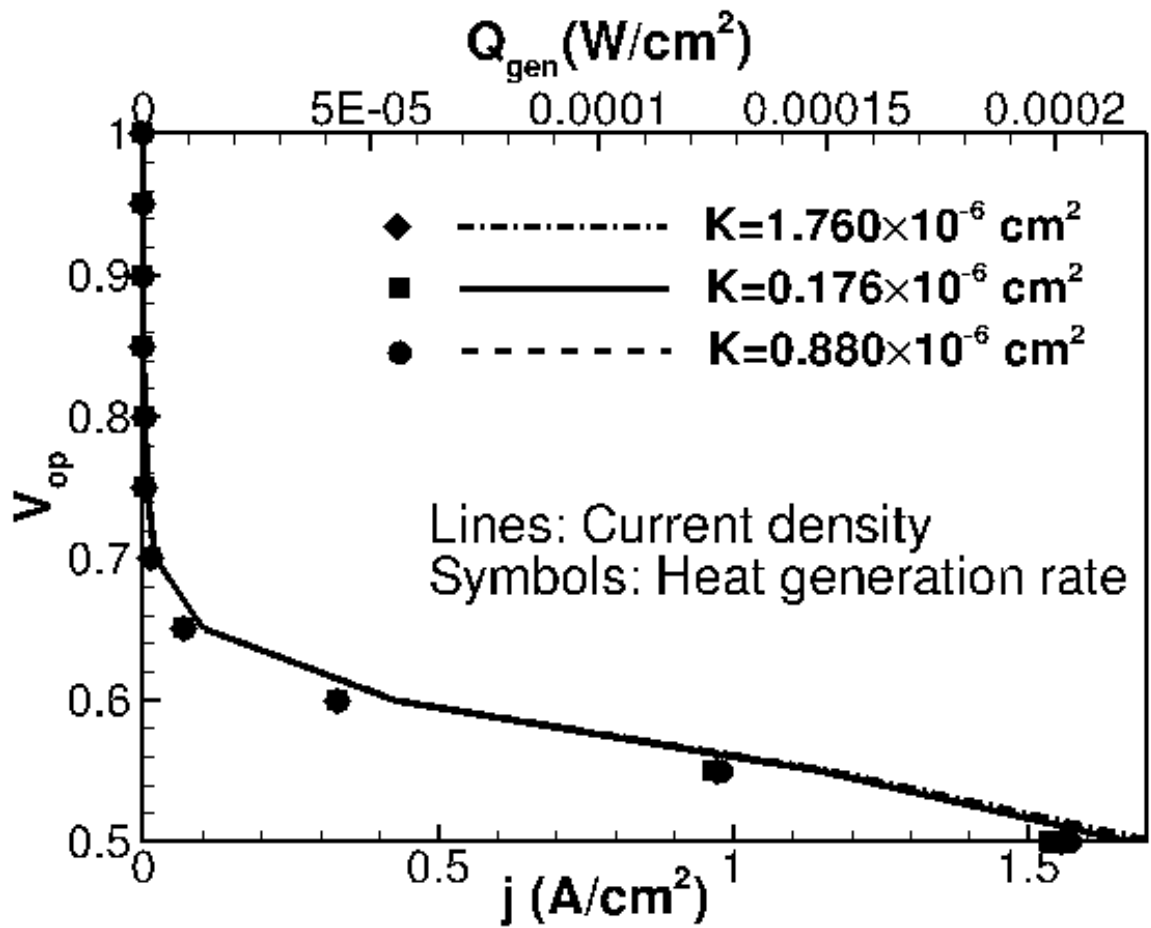

Figure 21: PERFORMANCE AND HEAT GENERATION RATE VARIATION WITH PERMEABILITY

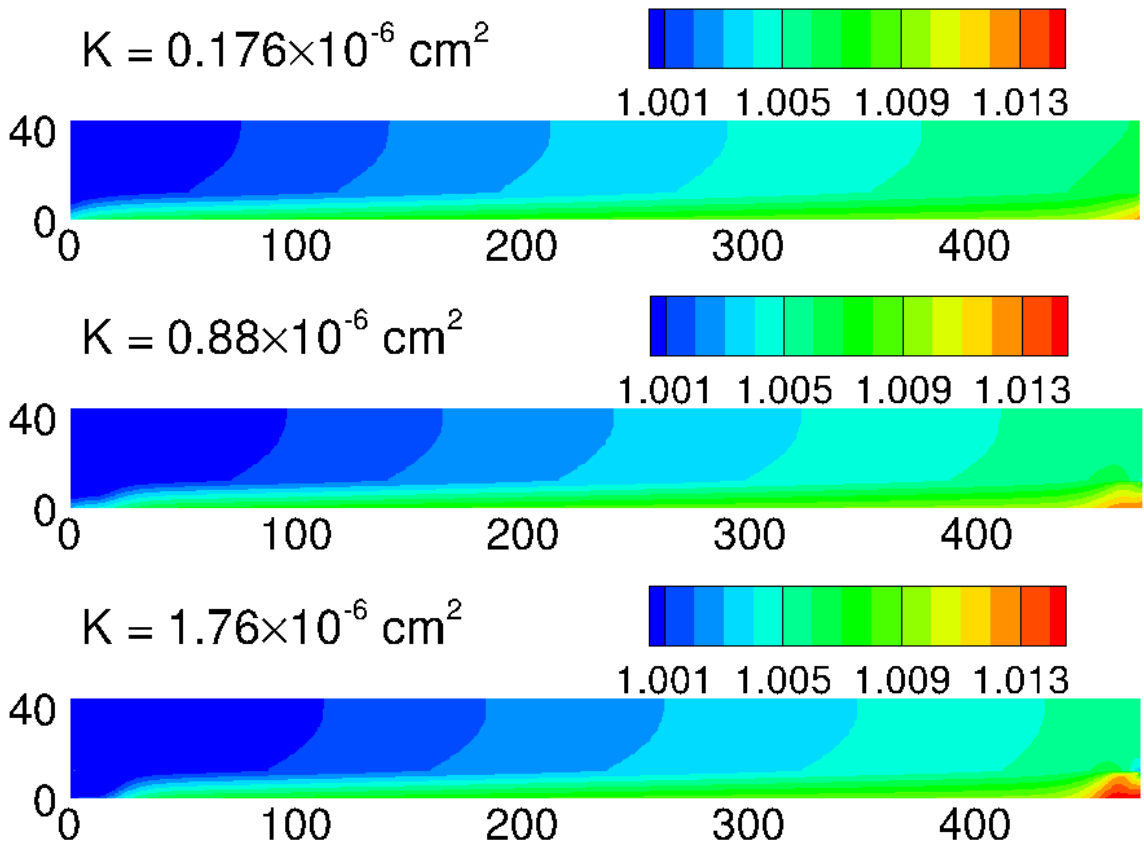

Figure 22: TEMPERATURE DISTRIBUTION FOR $V_{op}$-0.55V, FOR DIFFERENT PERMEABILITIES

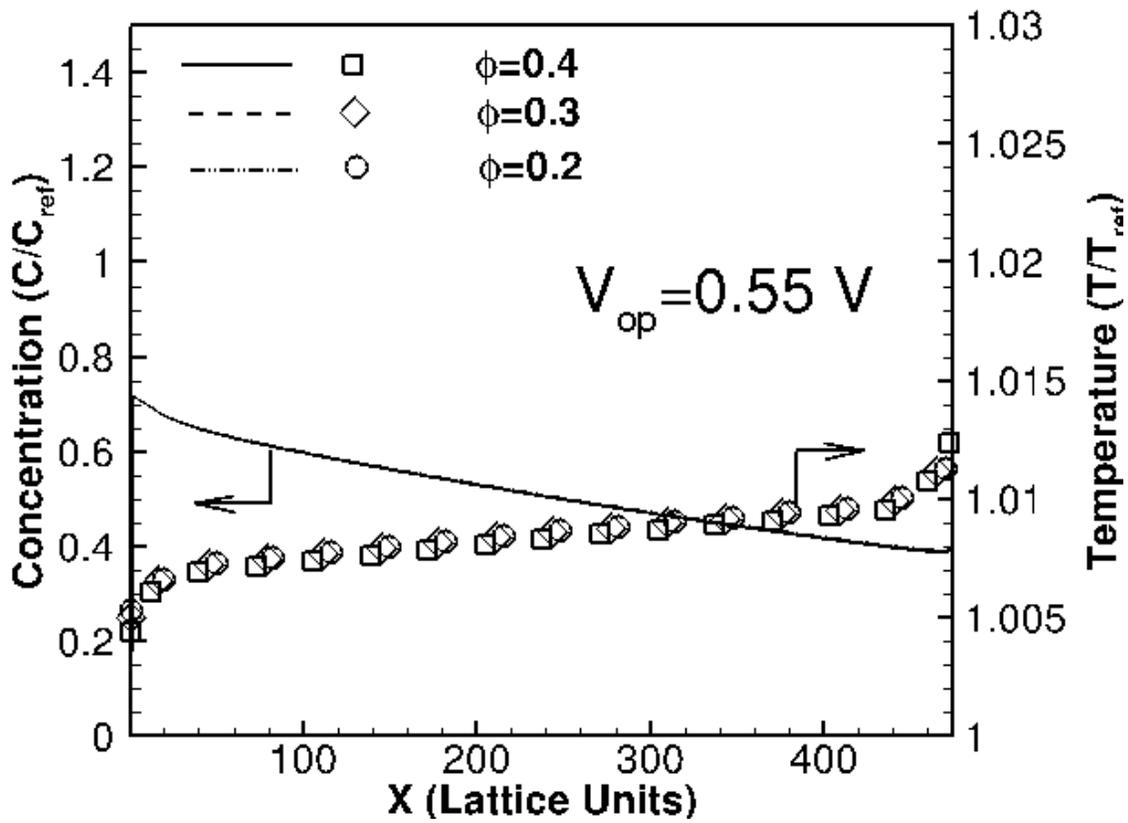

Figure 23: OXYGEN AND TEMPERATURE DISTRIBUTION ALONG CATALYST LAYER FOR DIFFERENT POROSITY OF GDL

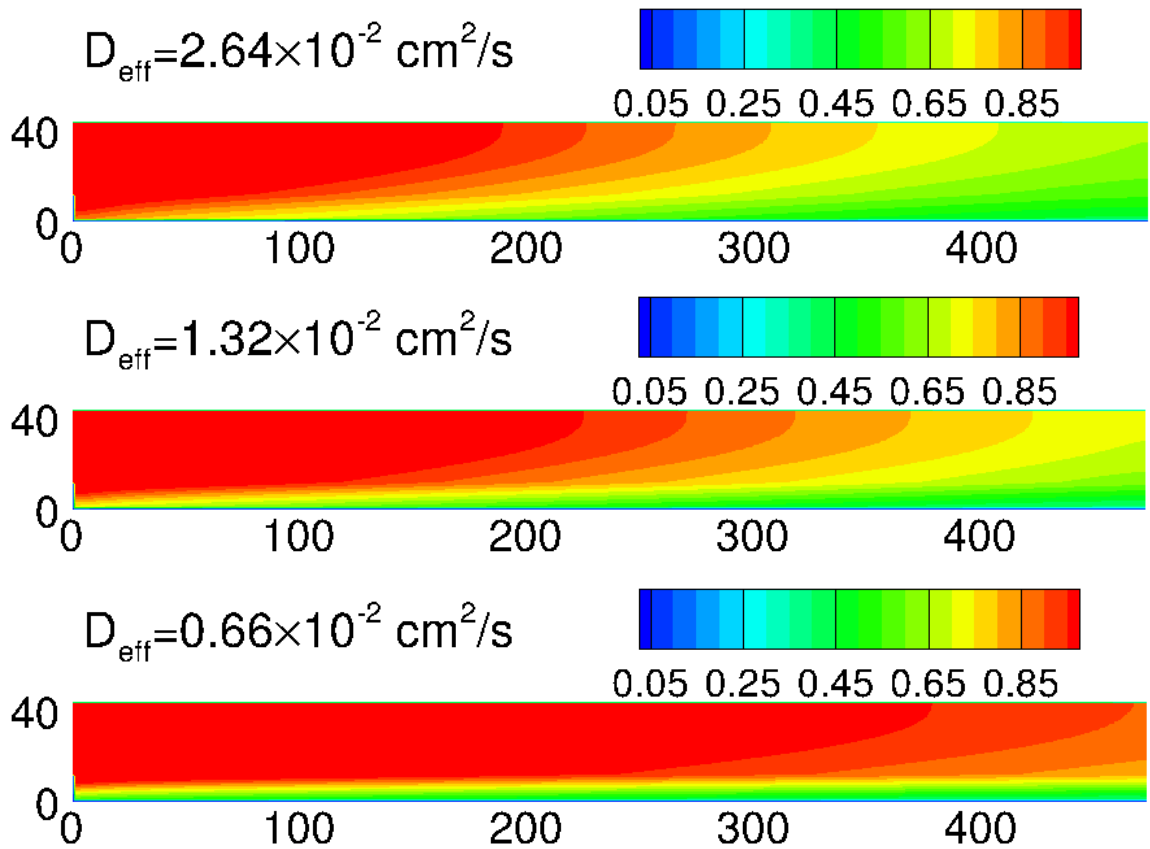

Figure 24: OXYGEN CONCENTRATION CONTOURS FOR DIFFERENT POROUS MEDIA EFFECTIVE DIFFUSIVITIES OF OXYGEN

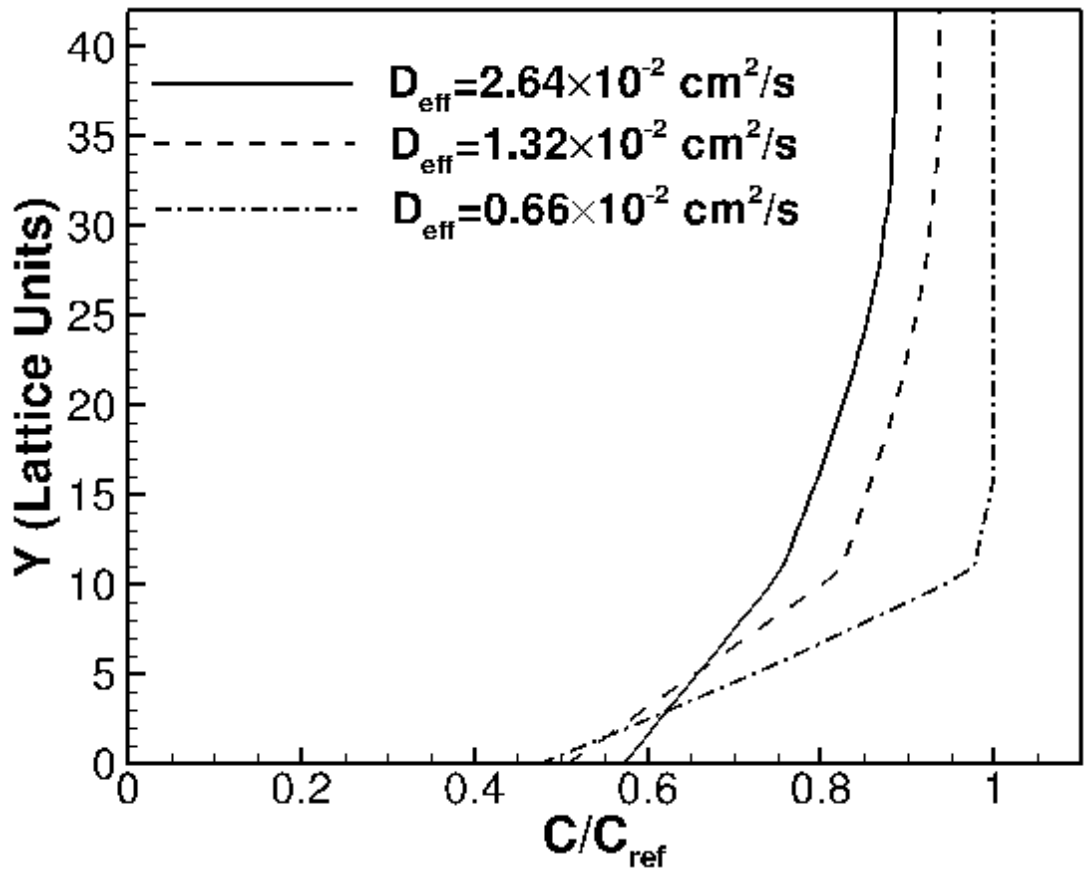

Figure 25: VARIATION OF OXYGEN CONCENTRATION IN THE DIRECTION NORMAL TO THE FLOW AT X=L/2.

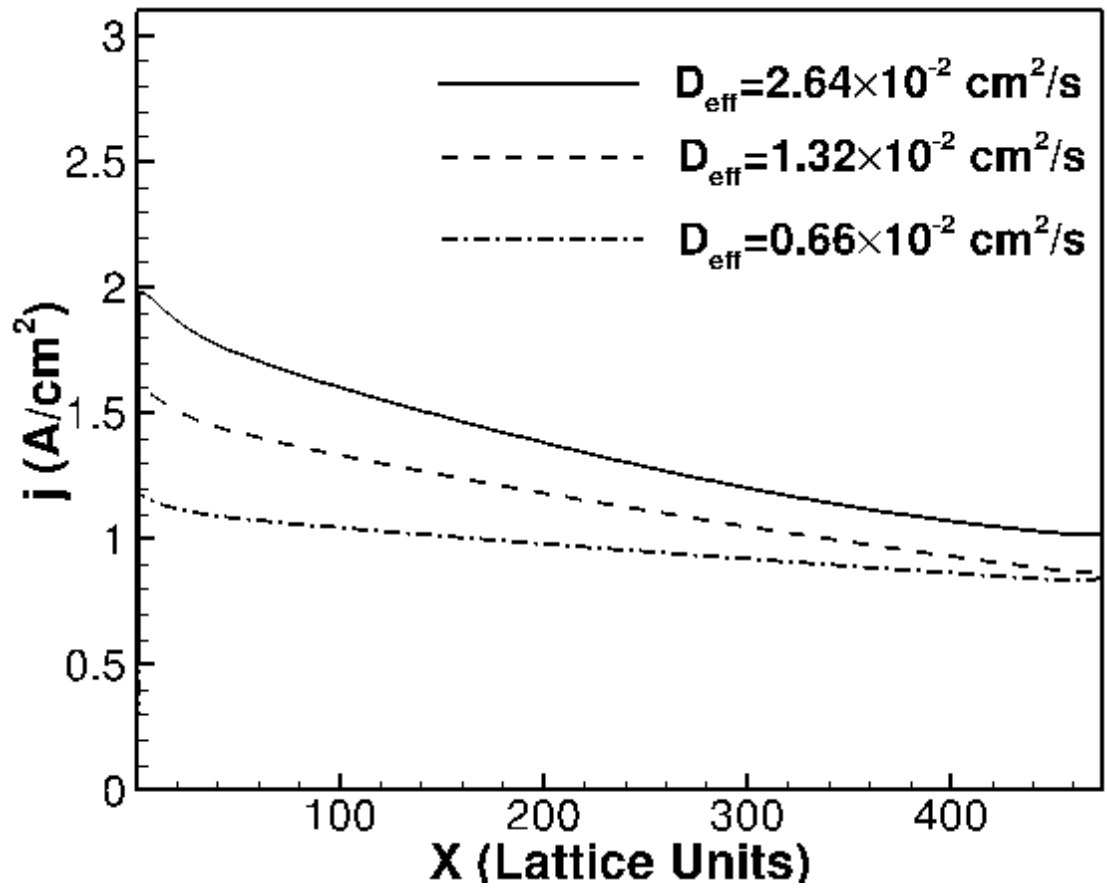

Figure 26: CURRENT DENSITY VARIATION ALONG CATALYST LAYER FOR DIFFERENT POROUS MEDIA EFFECTIVE OXYGEN DIFFUSIVITY VALUES

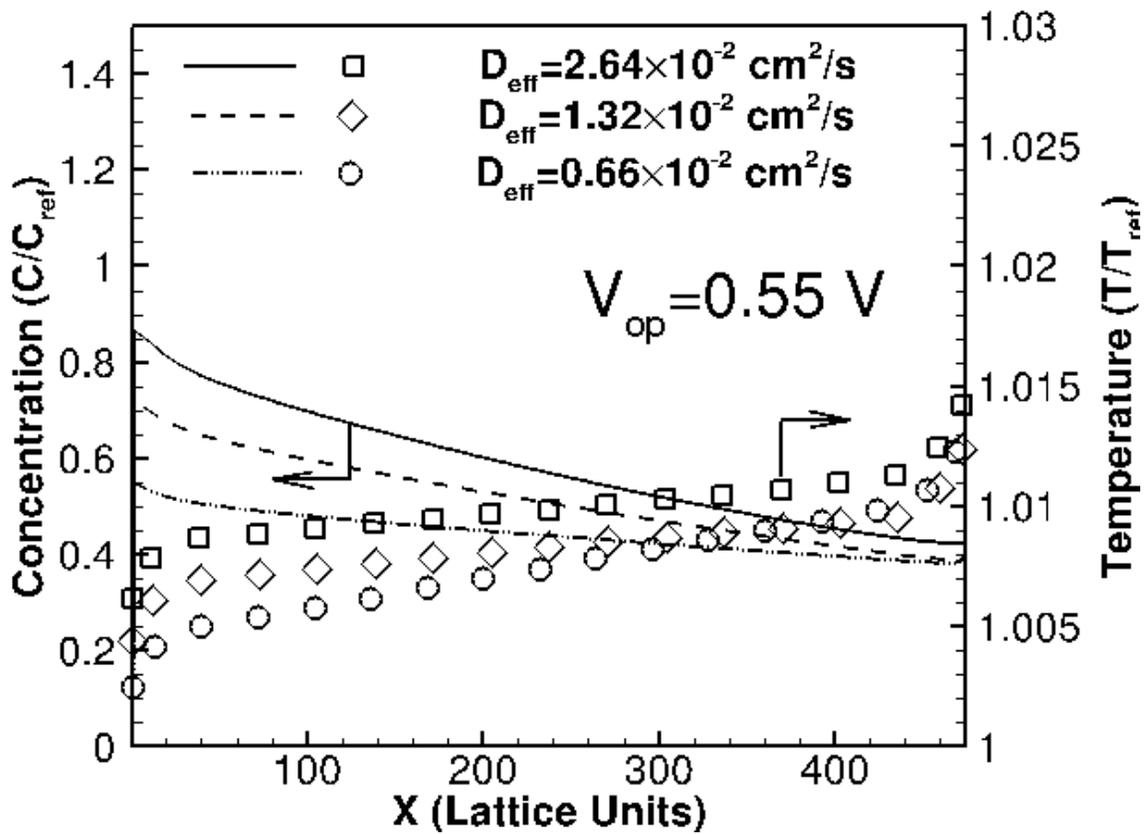

Figure 27: CONCENTRATION AND TEMPERAUTURE VARIATION ALONG CATALYST LAYER FOR DIFFERENT POROUS MEDIA EFFECTIVE OXYGEN DIFFUSIVITY VALUES

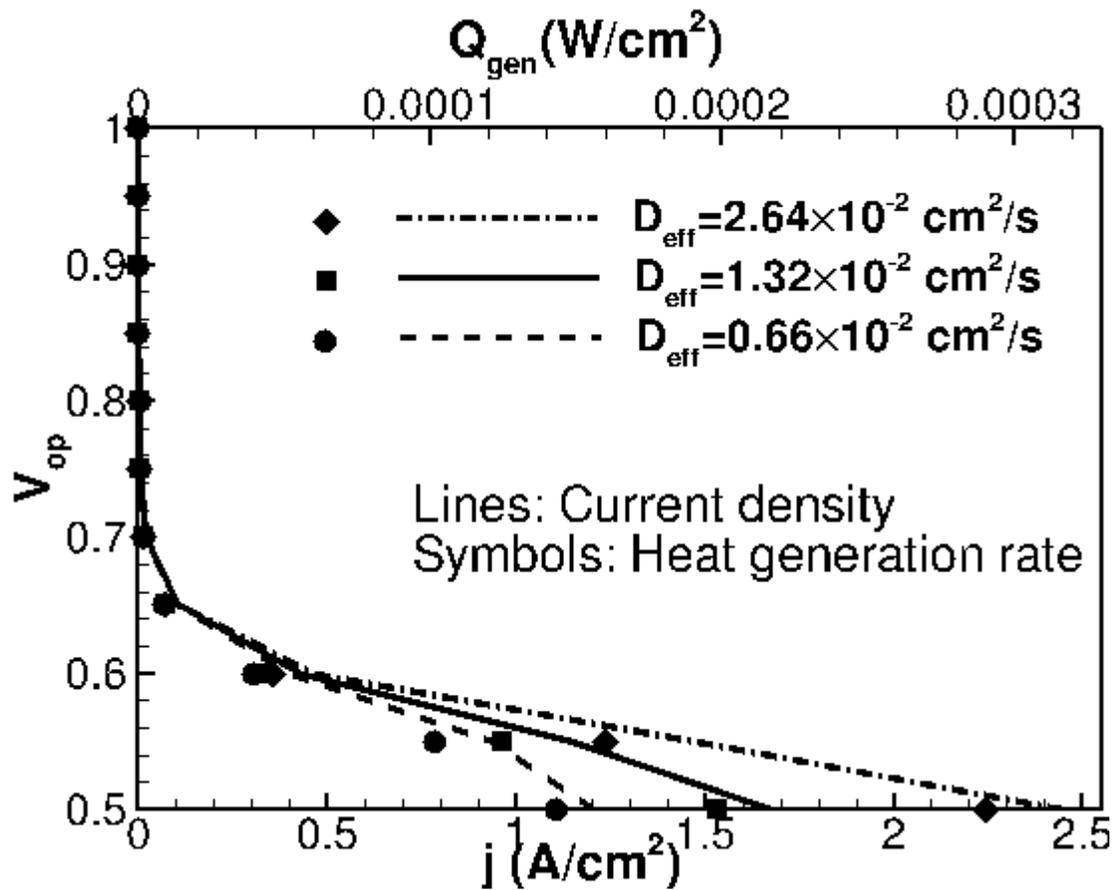

Figure 28: PERFORMANCE AND HEAT GENERATION RATE FOR DIFFERENT POROUS MEDIA EFFECTIVE OXYGEN DIFFUSIVITY VALUES

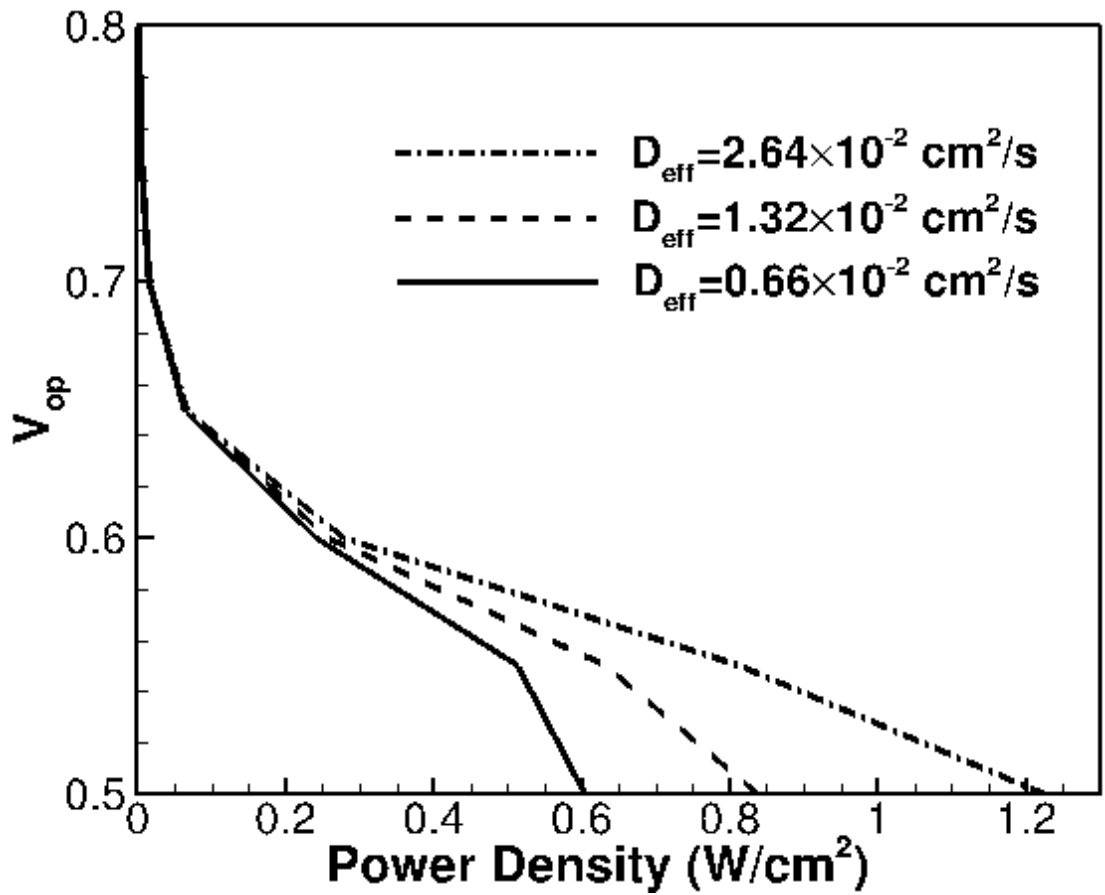

Figure 29: OUTPUT POWER DENSITY VARIATION FOR DIFFERENT POROUS MEDIA EFFECTIVE OXYGEN DIFFUSIVITY VALUES

[58]